\documentclass[nofootinbib,pra,twocolumn]{revtex4-2}
\usepackage{float,graphicx,multirow, amsmath,color,dsfont}
\usepackage{amsmath,bm}
\usepackage{enumerate}
\usepackage{amssymb,mathrsfs,dsfont}
\usepackage{amssymb,mathbbol}
\usepackage{amsfonts}
\usepackage{mathtools}
\usepackage{mathrsfs}
\usepackage{float}
\usepackage{xcolor,hyperref}
\definecolor{darkblue}{rgb}{0,0.0.1,0.3}
\definecolor{darkred}{rgb}{0.6,0.1,0}
\hypersetup{colorlinks,breaklinks,
	linkcolor=darkred,urlcolor=darkblue,
	anchorcolor=darkred,citecolor=
	darkred,pdfauthor=Chandan, pdftitle=ChandanDistillation}
\usepackage{tikz}
\usepackage{pifont}
\usepackage{mathbbol}
\usepackage{float}
\usepackage[normalem]{ulem} 

\newcommand{\ie}{\textit{i}.\textit{e}.}

\begin{document}
	
	\title{Advantage of probabilistic non-Gaussian operations in the distillation of single mode squeezed vacuum state   }
	\author{Chandan Kumar}
	\email{chandan.quantum@gmail.com}
	\affiliation{Optics and Quantum Information Group, The Institute of Mathematical Sciences, CIT Campus, Taramani, Chennai 600113, India.}
	\affiliation{Homi Bhabha National Institute, Training School Complex, Anushakti Nagar, Mumbai 400085, India.}
	
	\begin{abstract}
		We consider the distillation of squeezing in single mode squeezed vacuum   state using three different probabilistic non-Gaussian operations: photon subtraction (PS), photon addition (PA) and photon catalysis (PC). To accomplish this,  we consider a practical model to implement these non-Gaussian operations and derive the Wigner characteristic function of the resulting non-Gaussian states. Our result shows that while PS and PC operations can distill  squeezing, PA operations cannot. 
		Furthermore, we delve into the success probabilities associated with these non-Gaussian operations and identify optimal parameters for the distillation of squeezing. Our current analysis holds significant relevance for experimental endeavors concerned with squeezing distillation.

	\end{abstract}
	\maketitle

	\section{Introduction}
	
	Squeezing is an important quantum resource indispensable in various continuous variable quantum information processing  (QIP) protocols~\cite{Braunstein,Silberhorn,weedbrook-rmp-2012,Adesso}, encompassing applications like quantum teleportation~\cite{Furusawa,tele}, quantum metrology~\cite{metro,Giovannetti}, and quantum key distribution~\cite{qkd,qkd1,Pirandola:20}.
	While a maximum squeezing value of 15 dB has been demonstrated in the infrared wavelength regime~\cite{Vahlbruch, Mehmet:11, 15db},generating comparably intense squeezed states remains a formidable task in alternative wavelength ranges or different systems~\cite{Multistep}. For instance, squeezed states of magnitudes 1.1 dB~\cite{1.1dB} and 1.3 dB~\cite{1.3dB} were produced in optomechanical systems, portraying the challenges in producing higher levels of squeezing outside the infrared spectrum. Therefore, it is of paramount importance to find ways to enhance squeezing of the produced state.

	Gaussian operations and homodyne detection are ineffective in distilling  squeezing of a single mode Gaussian state~\cite{Kraus}. This is similar to the case of no-go theorem for entanglement distillation from two-mode Gaussian state via local Gaussian operations~\cite{Eisert,prl2002,Giedke}.
	Instead, we can resort to probabilistic non-Gaussian operations such as photon subtraction (PS) and  photon addition (PA)~\cite{Ourjoumtsev,Takahashi2010,Lvovsky,Dirmeier:20} for distillation of squeezing.
	Essentially, our aim is to prepare an ensemble of small size with high squeezing by selecting   few  from  an ensemble of large size heralded on the successful implementation of non-Gaussian operations.
	We note that non-Gaussian operations have been utilized to enhance the efficiency of different QIP  protocols, including  quantum
	metrology~\cite{gerryc-pra-2012,josab-2012,braun-pra-2014,josab-2016,pra-catalysis-2021,ill2008,ill2013,metro-thermal-arxiv, ngsvs-arxiv} and   quantum
	teleportation~\cite{tel2000,dellanno-2007,tel2009,catalysis15,catalysis17,wang2015,tele-2023,noisytele}. They have also been used to design quantum heat engine~\cite{Walmsley,Filip,Scarani,Zanin,Zhang,Buller}.

	In Ref.~\cite{Multistep}, multistep distillation scheme was employed to distill  squeezing from   an ensemble of   single mode squeezed vacuum   states (SVS).   The initial step involved executing two photon subtraction (2-PS) operations on the SVS state.
	The inquiry arises: Why was the 2-PS operation specifically chosen? Could a single photon subtraction (1-PS) operation or alternative non-Gaussian operations like   PA or photon catalysis (PC) have been utilized for squeezing distillation instead?

	To address these questions, we consider a practical model for implementing PS, PA and PC operations on SVS. The generated non-Gaussian states namely   photon subtracted SVS (PSSVS), photon added SVS (PASVS), and photon catalyzed SVS (PCSVS) will be collectively referred as NGSVSs. We derive the Wigner characteristic function of the NGSVSs, which is used to evaluate the quadrature squeezing. 
	The investigation into quadrature squeezing reveals distinct outcomes for different non-Gaussian operations as shown in Table~\ref{table}.
	
	\begin{table}[h!]
		\centering
		\caption{\label{table}
			Advantage (\ding{51}) and disadvantage (\ding{55}) of $n$-PS, $n$-PA, and $n$-PC operation in distillation of squeezing}
		\renewcommand{\arraystretch}{1.5}
		\begin{tabular}{ |c|c |c|c |c|c|c|}
			\hline \hline
			Operation & $n=1$  & $n=2$ & $n=3$ &$n=4$\\ \hline
			PS & \ding{55} & \ding{51} & \ding{55} & \ding{51}   \\\hline
			PA & \ding{55} & \ding{55} & \ding{55} & \ding{55}   \\\hline
			PC & \ding{55} & \ding{51} & \ding{51} & \ding{51}   \\\hline
		\end{tabular}
	\end{table}
	
	We note in particular that the application of two photon catalysis (2-PC) substantially amplifies the   squeezing  when the initial SVS state exhibits  small squeezing.

	In most of the theoretical analyses,  PS and PA operations is   implemented via the annihilation $\hat{a}$ and   creation $\hat{a}^\dagger$ operator, respectively.   However, we departed from this approach by considering a practical model to implement   non-Gaussian  operations, encompassing PS and PA operations. This model enables us to factor in the success probability associated with  non-Gaussian  operations, which directly correlates with resource utilization. 
	There are instances where quadrature squeezing can be maximized, yet the success probability is very close to zero. Such circumstances are impractical for experimental implementation. Instead, we propose a trade-off strategy, balancing the enhancement in quadrature squeezing against the success probability to attain optimal scenarios. We delineate the parameters corresponding to these optimal scenarios.

	The subsequent sections of this article are organized as follows.
	In Sec.~\ref{sec:wig}, we have derived the Wigner characteristic function of the SVS state. 
	In Sec.~\ref{sec:res}, we analyze the quadrature squeezing of the NGSVSs and also factor the success probability of the non-Gaussian operations.
	Finally, Sec.~\ref{sec:con} concludes the main results and provide future directions.
	\section{ Setup for the implementation of non-Gaussian operations }\label{sec:wig}

	\begin{figure}[H]
		\centering
		\includegraphics[scale=1]{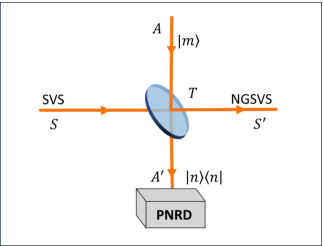}
		\caption{Setup for the implementation of non-Gaussian operations on the SVS state  (explained in detail in the text).  }
		\label{figsub}
	\end{figure}
	In order to implement different non-Gaussian operations on SVS, we consider the setup shown in Fig.~\ref{figsub}. The signal mode  (SVS)  is mixed with auxiliary mode   (Fock state) using a beam splitter of variable transmissivity. Photon number resolving detector is employed on the output mode and a successful detection heralds the implementation of non-Gaussian operation.

	We work in phase space and utilize the Wigner characteristic function formalism  for convenience in the calculation of moments of the quadrature operators appearing in the quadrature squeezing. Let us consider an   an $n$-mode quantum system with density operator $\hat{\rho}$, where  the $i$th mode has associated quadrature operators $\hat{q_i}$ and $\hat{p_i}$. The Wigner characteristic function of a state with density operator $\hat{\rho}$  is defined as~\cite{ olivares-2012,weedbrook-rmp-2012}, 
	\begin{equation}\label{wigde}
		\chi(\Lambda) = \text{Tr}[\hat{\rho} \, \exp(-i \Lambda^T \Omega \hat{\xi})],
	\end{equation}
	where $\hat{\xi} = (\hat{q_1}, \hat{p_1},\dots \hat{q_n}, \hat{p_n})^T$,  
	$\Lambda = (\Lambda_1, \Lambda_2, \dots \Lambda_n)^T$ with  
	$\Lambda_i = (\tau_i, \sigma_i)^T \in \mathcal{R}^2$. The symbol $\Omega$ denotes the symplectic form that is defined as,
	\begin{equation}
		\Omega = \bigoplus_{k=1}^{n}\omega =  \begin{pmatrix}
			\omega & & \\
			& \ddots& \\
			& & \omega
		\end{pmatrix}, \quad \omega = \begin{pmatrix}
			0& 1\\
			-1&0 
		\end{pmatrix}.
	\end{equation}

	For Gaussian states, we require only the first and second order moments. First order moments or the mean displacement vector $\overline{\xi}$ can be written as   
	\begin{equation}
		\overline{\xi} = \langle  \hat{\xi } \rangle =
		\text{Tr}[\hat{\rho} \hat{\xi}].
	\end{equation} 
	The second order moments can be expressed in the form of a matrix known as covariance matrix   $V$, which is  defined as
	\begin{equation}\label{eq:cov}
		V = (V_{ij})=\frac{1}{2}\langle \{\Delta \hat{\xi}_i,\Delta
		\hat{\xi}_j\} \rangle,
	\end{equation}
	where $\Delta \hat{\xi}_i = \hat{\xi}_i-\langle \hat{\xi}_i
	\rangle$, and $\{\,, \, \}$ represents anti-commutator.
	The corresponding Wigner characteristic function is easily obtainable from the following expression~\cite{weedbrook-rmp-2012}:
	\begin{equation}\label{wig:g}
		\chi(\Lambda) =\exp[-\frac{1}{2}\Lambda^T (\Omega V \Omega^T) \Lambda- i (\Omega\overline{\xi} )^T\Lambda].
	\end{equation}

	We represent the associated quadrature operators of the system mode   with ($\hat{q_1}$,$\hat{p_1}$) and that of the auxiliary mode   with ($\hat{q_2}$,$\hat{p_2}$). 
	For the vacuum state, which has the covariance matrix $V=$ diag$(1/2,1/2)$, the Wigner characteristic function~(\ref{wig:g}) reads 
	\begin{equation}
		\chi_{|0\rangle} (\Lambda_1)= \exp\left [   -\left( \tau_ 1^2 +  \sigma_ 1^2\right)/4 \right].
	\end{equation} 
	Under the squeezing transformation characterized by the transformation matrix $S=$ diag$(e^{-r},e^r)$, the Wigner characteristic function transforms as $\chi(\Lambda) \rightarrow \chi\left(S^{-1} \Lambda\right)$~\cite{weedbrook-rmp-2012}. Thus, the Wigner characteristic function of the SVS turns out to be 
	\begin{equation}
		\chi_\text{svs} (\Lambda_1)= \exp\left [   -\left(e^{2 r}\tau_ 1^2 + e^{-2 r}\sigma_ 1^2\right)/4 \right].
	\end{equation} 
	Prior to the beam splitter operation, the Wigner characteristic function of the two-mode system is given by 
	\begin{equation}
		\chi_{\text {svs}, |m\rangle}(\Lambda)=\chi_{\text {svs}}\left(\Lambda_1\right) \chi_{|m\rangle}\left(\Lambda_2\right).
	\end{equation}
	Here $\chi_{|m\rangle}\left(\Lambda_2\right)$ is the Wigner characteristic function of the Fock state $|m\rangle$, which can be evaluated using Eq.~(\ref{wigde}):
	\begin{equation}\label{charfock}
		\chi_{|m\rangle}(\Lambda_2)=\exp  \left[- \frac{\tau_2^2}{4}-\frac{\sigma_2^2}{4} \right]\,L_{m}\left( \frac{\tau_2^2}{2}+\frac{\sigma_2^2}{2} \right),
	\end{equation}
	where $L_m(\bullet)$ is the Laguerre polynomial.
	The  transformation matrix corresponding to the beam splitter operation is given by
	\begin{equation}\label{beamsplitter}
		B(T) = \begin{pmatrix}
			\sqrt{T} \,\mathbb{1}_2& \sqrt{1-T} \,\mathbb{1}_2 \\
			-\sqrt{1-T} \,\mathbb{1}_2& \sqrt{T} \,\mathbb{1}_2
		\end{pmatrix},
	\end{equation}
	where $\mathbb{1}_2$ represents the $2 \times 2$ identity matrix. 
	Post the beam splitter operation, the Wigner characteristic function of the output entangled state can be evaluated as 
	\begin{equation}
		\chi_{\text {out}}(\Lambda)=\chi_{\text {svs}, |m\rangle}(B(T)^{-1}\Lambda).
	\end{equation}
	A photon number resolving detector   measuring $n$ photons in the output auxiliary mode leads to the generation of NGSVSs. The corresponding   Wigner characteristic function is given by
	\begin{equation}\label{unnorm}
		\chi_{\text {NGSVSs}}\left(\Lambda_1\right)=\frac{1}{2 \pi} \int d \Lambda_1 d \Lambda_2 \chi_{\text {out }}(\Lambda) \chi_{|n\rangle}\left(\Lambda_2\right).
	\end{equation}
	We can write the Fock state using the following identity as
	\begin{equation}\label{charfock1}
		\chi_{|k\rangle}(\tau_i,\sigma_i)=\exp  \left[- \frac{\tau_i^2}{4}-\frac{\sigma_i^2}{4} \right]\,	\bm{\widehat{F}}e^{ 2 uv +u(\tau_i+i\sigma_i)-v(\tau_i-i\sigma_i)},
	\end{equation}
	with
	\begin{equation}
		\bm{\widehat{F}} =  \frac{1}{2^k k!}  \frac{\partial^k}{\partial\,u^k} \frac{\partial^k}{\partial\,v^k} \{ \bullet \}_{u=v=0}.
	\end{equation}
	This enables us to convert the integrand of Eq.~(\ref{unnorm}) into a Gaussian function, which can be easily evaluated. The integral turns out to be
	\begin{equation}\label{unnormwig}
		\chi_{\text{NGSVSs}}\left(\Lambda_1\right) =a_0 \bm{\widehat{D} } \exp 
		\left(\bm{\Lambda}^T G_1 \bm{\Lambda}+\bm{u}^T G_2 \bm{\Lambda} + \bm{u}^T G_3 \bm{u} \right),
	\end{equation}
	where $a_0 = \sqrt{(1-\lambda^2)/(1-\lambda^2 T^2)}$ with $\lambda=\tanh r$. The column vectors are given by $\bm{\Lambda}=(\tau_1,\sigma_1)^T$ and $\bm{u}=(u_1,v_1,u_2,v_2)^T$. The matrices are: 
	\begin{equation}
		G_1= \frac{a_0 }{-4(1-\lambda^2 T^2)} \begin{pmatrix}
			(1+\lambda T)^2 &0\\
			0 &(1-\lambda T)^2
		\end{pmatrix},
	\end{equation}

	\begin{equation}
		G_2= \frac{a_0\sqrt{1-T}}{1-\lambda^2 T^2} \left(
		\begin{array}{cc}
			\lambda  T+1 & -i (\lambda  T-1) \\
			-\lambda   T-1 & -i (\lambda  T-1) \\
			-\lambda  \sqrt{T} (\lambda  T+1) & -i \lambda  \sqrt{T} (\lambda  T-1) \\
			\lambda  \sqrt{T} (\lambda  T+1) & -i \lambda  \sqrt{T} (\lambda  T-1) \\
		\end{array}
		\right),
	\end{equation}
	
	and

	\begin{widetext}
		\begin{equation}
			G_3= \frac{a_0 }{1-\lambda^2 T^2} \left(
			\begin{array}{cccc}
				\lambda T (T-1)  & 1-T &   \lambda \sqrt{T} (1-T)   & \sqrt{T} \left(1-\lambda ^2 T\right) \\
				1-T & \lambda T (T-1)  & \sqrt{T} \left(1-\lambda ^2 T\right) & 
				\lambda \sqrt{T} (1-T)   \\
				\lambda \sqrt{T} (1-T)   & \sqrt{T} \left(1-\lambda ^2 T\right) & \lambda  (T-1) &
				\lambda^2 T(1-T)   \\
				\sqrt{T} \left(1-\lambda ^2 T\right) &   \lambda \sqrt{T}  (1-T)  &   \lambda ^2 T (1-T )   & \lambda  (T-1) \\
			\end{array}
			\right).
		\end{equation}
	\end{widetext}
	The operator $\bm{\widehat{D} } $ is defined as 
	\begin{equation}
		\begin{aligned}
			\bm{\widehat{D} } = \frac{2^{-(m+n)}}{m!n!} \frac{\partial^{m}}{\partial\,u_1^{m}} \frac{\partial^{m}}{\partial\,v_1^{m}} \frac{\partial^{n}}{\partial\,u_2^{n}} \frac{\partial^{n}}{\partial\,v_2^{n}}
			\{ \bullet \}_{\substack{u_1= v_1= 0\\ u_2= v_2= 0}}.\\
		\end{aligned}
	\end{equation}

	We note that the above Wigner characteristic function is unnormalized~(\ref{unnormwig}). The normalized Wigner characteristic function is given by
	\begin{equation}
		\widetilde{\chi}_{\text{NGSVSs}}= P^{-1}\chi_{\text{NGSVSs}}.
	\end{equation}
	
	Here $P$ is the probability given by
	\begin{equation}\label{probng}
		\begin{aligned}
			P &=\chi_{\text{NGSVSs}}\bigg|_{\tau_1= \sigma_1=0}
			= \bm{\widehat{D} } \exp 
			\left(\bm{u}^T G_3 \bm{u} \right). \\		
		\end{aligned}
	\end{equation}
	By choosing specific combinations of $m$ and $n$, we can perform PS, PA and PC operations. When $m<n$ and $m>n$, the SVS undergoes PS and PA operations, respectively, resulting in states labeled as PSSVS and PASVS. In this article, we set $m=0$ and $n=0$ for PS and PA operations, respectively.  As the transmissivity approaches unity, the PS and PA operations become  ideal PS ($\hat{a}_1^n$) and ideal PA ($\hat{a}{^\dagger_1}^m$) operations. Surprisingly,  we observed that the Wigner characteristic function of the 1-PSSVS and 1-PASVS are identical. When $m=n$ is chosen, the SVS undergoes PC operation, yielding a state labeled as PCSVS. In the limit of unit transmissivity, the PC operation becomes identity and the PCSVS becomes the SVS.

	The Wigner characteristic function of NGSVSs can be differentiated with respect to $\tau_1$ and $\sigma_1$ to obtain the average of symmetrically ordered operators:
	
	\begin{equation}\label{app:covfinalch}
		\begin{aligned}
			{}_{\bm{:}}^{\bm{:}}  \hat{q_1}^{s } \hat{p_1}^{t }
			{}_{\bm{:}}^{\bm{:}}   =\left( \frac{1}{i} \right)^{s }\left( \frac{1}{-i} \right)^{t }
			\frac{\partial^{ s+t }}{\partial \sigma_1^{s } \partial \tau_1^{t } } 
			\{ 	\widetilde{\chi}_{\text{NGSVSs}}  \}_{\substack{\tau_1=
					\sigma_1=0  }}, 
		\end{aligned}
	\end{equation}
	where
	${}_{\bm{:}}^{\bm{:}}  \bullet {}_{\bm{:}}^{\bm{:}} $  is the notation for symmetric ordering~\cite{Barnett}.

	\section{Distillation of squeezing   }\label{sec:res}
	Having evaluated the Wigner characteristic function, we now move to the analysis of quadrature squeezing. The quadrature squeezing in the $\hat{q}$-quadrature given by
	\begin{equation}
		(\Delta q_1 )^2 = \langle \hat{q}_1^2 \rangle - \langle \hat{q}_1 \rangle ^2.
	\end{equation} 
	Since $\hat{q}_1$ and $\hat{q}_1^2$ are symmetric ordered operators, we  can choose appropriate values of ($s$, $t$)  
	in Eq.~(\ref{app:covfinalch}) to calculate  the quadrature squeezing.
	For the SVS,
	\begin{equation}\label{covsvs}
		(\Delta q_1 )^2_\text{SVS}=  \exp(-2r) =  -\frac{1}{2}+ \frac{1}{\lambda +1}.
	\end{equation} 
	Our aim is to analyze quadrature squeezing for the NGSVSs, seeking answers to previously posed questions. Specifically, we aim to identify the non-Gaussian operations capable of distilling squeezing.
	
	\subsection{Quadrature squeezing of 2-PSSVS}
	We first analyze the case of two photon subtraction ($m=0,n=2$), which was utilized in the first step of the multistep distillation scheme~\cite{Multistep}. For the 2-PSSVS,
	\begin{equation}\label{gensq}
		(\Delta q_1)^2_\text{2-PS}=  -\frac{5}{2} +\frac{5}{\lambda  T +1} +\frac{2 (\lambda  T -1)}{2 \lambda ^2 T ^2+1}.
	\end{equation}
	
	\begin{figure}[h!]
		\includegraphics[scale=1]{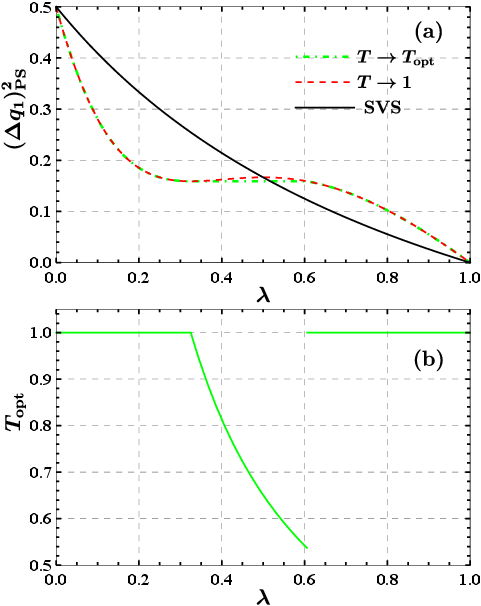}
		\caption{(a) Quadrature squeezing of 2-PSSVS as a function of squeezing parameter $\lambda$. (b) Optimal transmissivity minimizing the quadrature squeezing of 2-PSSVS~(\ref{gensq}).}
		\label{tauopt}
	\end{figure}

	In the unit transmissivity limit, the state is ideal 2-PSSVS \ie, $\mathcal{N}_s \hat{a}^2|\text{SVS}\rangle$, which has been analyzed theoretically in Ref.~\cite{Multistep}. The corresponding  quadrature squeezing can be obtained by taking the unit transmissivity limit in Eq.~(\ref{gensq}):
	\begin{equation}
		(\Delta q_1)^2_{T \rightarrow 1}= -\frac{5}{2}+  \frac{5}{\lambda +1}+ \frac{2 (\lambda -1)}{2 \lambda ^2+1}.
	\end{equation} 
	For   unit transmissivity, distillation of squeezing,    $\left[ (\Delta q_1)^2_{T \rightarrow 1} < (\Delta q_1)^2_\text{SVS} \right]$ is possible when $\lambda <1/2$~\cite{Multistep}.
	For variable transmissivity, 
	the quadrature squeezing~(\ref{gensq}) is minimized in the unit transmissivity limit in the range $\lambda \in \mathcal{S}_1 \equiv (0,0.33) \, {\displaystyle \cup } \, (0.61,1) $.
	Therefore, $(\Delta q_1)^2_{T \rightarrow 1}$ is the minimum value in the region $\mathcal{S}_1$. 
	In the range $\lambda \in \mathcal{S}_2 \equiv (0.33,0.61) $,  squeezing is minimized at
	\begin{equation}
		T_\text{opt}=\frac{-1 - \left (\frac {25} { 7+3 \sqrt{6}} \right)^{\frac{1}{3}} + \left (35+15 \sqrt{6} \right)^{\frac{1}{3}} }{6\lambda}   .
	\end{equation}
	The corresponding  minimized squeezing turns out to be
	\begin{equation}
		(\Delta q_1)^2_{T \rightarrow  T_\text{opt}} =\frac{1-5 \lambda  +14 \lambda ^2-10 \lambda ^3}{2 \left(1+\lambda   +2 \lambda ^2+ \lambda ^3\right)}.
	\end{equation}

	We have pictorially depicted these findings in Fig.~\ref{tauopt}. It is apparent that the 2-PS operation has the ability to extract squeezing. While quadrature squeezing is not optimized in the unit transmissivity limit in the region $\mathcal{S}_2$, the difference between $(\Delta q_1)^2_{T \rightarrow  T_\text{opt}}$ and $(\Delta q_1)^2_{T \rightarrow  1}$ is negligible.

	\subsection{Quadrature squeezing of NGSVSs}
	In Ref.~\cite{Multistep}, 2-PS operation was implemented to distill  squeezing in the first step of a multistep distillation scheme. Our investigation aims to assess alternative non-Gaussian operations that could potentially substitute the 2-PS operation in  such a distillation scheme.

	\begin{figure}[h!]
		\includegraphics[scale=1]{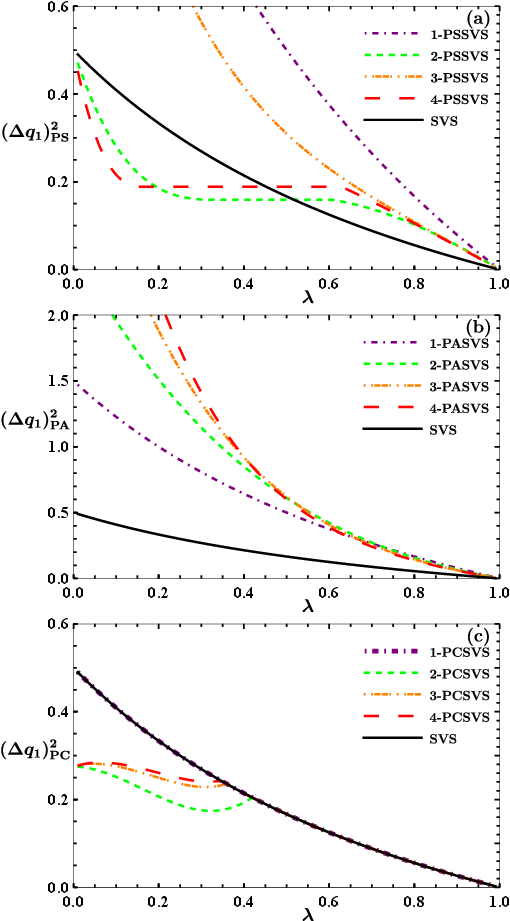}
		\caption{Quadrature squeezing $(\Delta q_1)^2$ optimized over the transmissivity as a function of squeezing $\lambda$ for NGSVSs.}
		\label{squeezing_1d}
	\end{figure}

	We plot the quadrature squeezing optimized over the transmissivity $T$ as a function of squeezing $\lambda$ for different NGSVSs in Fig.~\ref{squeezing_1d}.
	The 2-PS and 4-PS operations demonstrate an ability to enhance the quadrature squeezing of the SVS. However, the 1-PS and 3-PS operations result in a reduction of the quadrature squeezing of the SVS. 
	The quadrature squeezing for the 1-PSSVS is 
	\begin{equation} 
		(\Delta q_1 )^2_\text{1-PS}=   -\frac{3}{2}+ \frac{3}{1+\lambda T}.
	\end{equation}
	This is minimized in the unit transmissivity limit and $(\Delta q_1 )^2_\text{1-PS}\geq(\Delta q_1 )^2_\text{SVS}$ holds for   $\lambda \in [0,1)$. Therefore, 1-PS cannot distill  squeezing.  In the zero squeezing and unit transmissivity limit, the quadrature squeezing  is
	\begin{equation}
		\lim \limits_{\substack{%
				\lambda \to 0\\
				T \to 1}}	(\Delta q_1 )^2_\text{1-PS} =	\lim \limits_{\substack{%
				\lambda \to 0\\
				T \to 1}}	(\Delta q_1 )^2_\text{3-PS} =\frac{3}{2}.
	\end{equation}

	As we noted earlier that 1-PSSVS and 1-PASVS have the same Wigner characteristic function, therefore, 1-PA operation like 1-PS  operation   cannot distill  squeezing.  The quadrature squeezing for the 2-PASVS is 
	\begin{equation} 
		(\Delta q_1 )^2_\text{2-PA}=  -\frac{1}{2} +\frac{5}{\lambda  T +1} -\frac{2 (2+ \lambda  T  )}{2 + \lambda ^2 T ^2 }. 
	\end{equation}
	This is also minimized in the unit transmissivity limit and $(\Delta q_1 )^2_\text{2-PA}\geq(\Delta q_1 )^2_\text{SVS}$ holds for    $\lambda \in [0,1)$. Therefore, 2-PA lacks the ability to distill  squeezing. 
	Moreover, the observation from Fig.~\ref{squeezing_1d}(b) reveals that adding more photons  also  fails to  distill squeezing.  In the zero squeezing and unit transmissivity limit, the quadrature squeezing  is
	\begin{equation}
		\lim \limits_{\substack{%
				\lambda \to 0\\
				T \to 1}}	(\Delta q_1 )^2_\text{2-PA}  	  =\frac{5}{2},\quad	\lim \limits_{\substack{%
				\lambda \to 0\\
				T \to 1}}	(\Delta q_1 )^2_\text{3-PA}  	  =\frac{7}{2}.
	\end{equation}
	Moving to 1-PC operation, the corresponding quadrature squeezing of the 1-PCSVS state is given by 
	\begin{equation}
		(\Delta q_1)^2 	= 
		\left[\dfrac{\splitdfrac{ ( 1 - \lambda T)(1 + 4\lambda + 10\lambda^2 - 4\lambda T - 22\lambda^2T}
			{  -4\lambda^3T + 10\lambda^2T^2 + 
				4\lambda^3T^2 + \lambda^4T^2) }}
		{2 ( 1 + \lambda T)\left (1 + 2\lambda^2 - 6\lambda^2T + 
			2\lambda^2T^2 + \lambda^4T^2 \right)}\right].
	\end{equation}
	
	Quadrature squeezing reaches its minimum in the unit transmissivity limit, where $(\Delta q_1)^2_\text{1-PC}=(\Delta q_1)^2_\text{SVS}$. Consequently, the 1-PC operation lacks the capacity to distill squeezing. This is in contrast with the fact that 1-PC operation can distill squeezing from mixed state~\cite{catalysisprl2002,catalysisnature2015}.
	
	On the other hand, the 2-PC, 3-PC, and 4-PC operations demonstrate the ability to distill squeezing. These PC operations notably enhance quadrature squeezing by a substantial margin, especially for smaller initial squeezing parameters $\lambda$. Among the considered PC operations, the 2-PC operation exhibits the most promising performance.

\begin{figure}[h!]
		\includegraphics[scale=1]{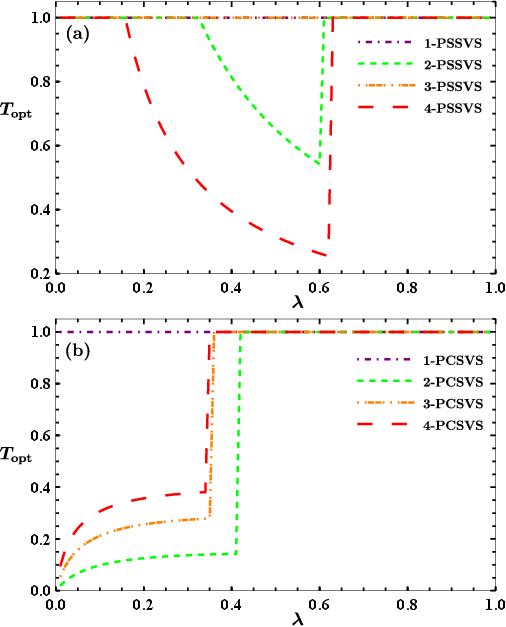}
		\caption{Optimal beam splitter transmisivity, minimizing the quadrature squeezing $(\Delta q_1)^2$, as a function of squeezing $\lambda$ for NGSVSs.}
		\label{tau_1d}
	\end{figure}

	We have presented the optimal beam splitter transmissivity minimizing the quadrature squeezing as a function of $\lambda$ in Fig.~\ref{tau_1d}.
	For PA operation, $T_\text{opt}$ turns out to be unity.

	It is enlightening to identify  the specific range of squeezing and transmissivity parameter rendering distillation of squeezing. To achieve this, we study the enhancement in quadrature squeezing  of the generated NGSVSs in two parameter space namely squeezing and transmissivity. This enhancement, denoted by $\mathcal{D}_\text{NG}$, is defined as following:
	\begin{equation}
		\mathcal{D}_\text{NG}= {(\Delta q_1)^2}_\text{SVS}- {(\Delta q_1)^2}_\text{NGSVSs}.
	\end{equation}
	
\begin{figure}[h!]
		\includegraphics[scale=1]{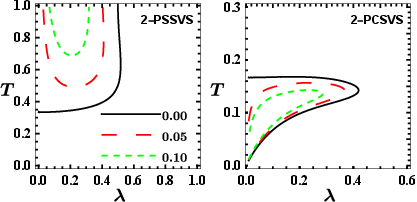}
		\caption{Curves of fixed $\mathcal{D}_\text{NG}$, the difference of quadrature squeezing between  SVS and NGSVSs, as a function of squeezing $\lambda$ and transmissivity $T$.}
		\label{contour}
	\end{figure}

	Illustrated in Fig.~\ref{contour} are the plots of $\mathcal{D}_\text{NG}$ for both 2-PSSVS and 2-PCSVS. In the case of 2-PSSVS, squeezing distillation is viable for low squeezing and high transmissivity values. Conversely, for 2-PCSVS, squeezing can be extracted within the realm of low squeezing and low transmissivity values.
	
	The area within the $r$ and $T$   parameter space, where $\mathcal{D} ^{\text{NG} }$ becomes positive, signifies the potential of non-Gaussian operations to distill squeezing.
	
	We have illustrated definite valued curves of the enhancement $\mathcal{D}_\text{NG}$ for both 2-PSSVS and 2-PCSVS  in Fig.~\ref{contour}  In the case of 2-PSSVS, the distillation of squeezing is feasible for low squeezing and high transmissivity values. For 2-PCSVS, squeezing can be distilled for low squeezing and low transmissivity values.
	
	\subsection{Success probability and optimal parameters}
	
\begin{figure}[h!]
		\includegraphics[scale=1]{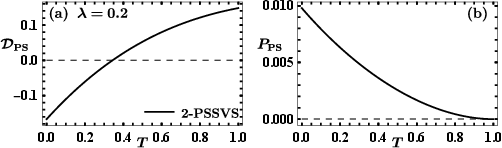}
		\caption{ (a) Enhancement in quadrature squeezing of the original SVS and 2-PSSVS, $\mathcal{D}_\text{NG}$, as a function of transmissivity $T$. (b) Success probability as a function of transmissivity $T$ for 2-PSSVS.  }
		\label{prob_1d}
	\end{figure}

	As depicted in Fig.~\ref{figsub}, the implementation of non-Gaussian operations relies on the   detection of a definite photon number by the photon number resolving detector. This introduces a probabilistic nature to the execution of non-Gaussian operations. Consequently, when selecting optimal parameters for a distillation experiment, it is crucial to consider the success probability. To emphasize this aspect, we focus on the 2-PSSVS and illustrate the relationship between the enhancement in quadrature squeezing $\mathcal{D}_\text{NG}$ and the success probability plotted against transmissivity $T$ in Fig.~\ref{prob_1d}.

	The analytical expression for the success probability for the 2-PSSVS is given by
	\begin{equation}
		P = \frac {\lambda^2 (1 - T)^2\sqrt {1 - \lambda^2} (1 + 
			2\lambda^2 T^2)} {4 (1 - \lambda^2 T^2)^{5/2}}.
	\end{equation}
	The graph illustrates that while the enhancement $\mathcal{D}_\text{NG}$ is maximized in the unit transmissivity limit, the probability tends toward zero under the same conditions.  
	Therefore, aiming to maximize the enhancement $\mathcal{D}\text{NG}$ results in a scenario unsuitable from an experimental perspective.
	
	\begin{figure}[h!]
		\includegraphics[scale=1]{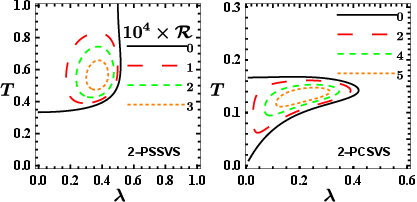}
		\caption{Curves of fixed magnitude of the product $\mathcal{R} = \mathcal{D}_\text{NG} \times P$   as a function of squeezing $\lambda$ and transmissivity $T$.  }
		\label{prob_contour}
	\end{figure}
	To obtain an optimal scenario, we strike a balance between the enhancement in quadrature squeezing $\mathcal{D}\text{NG}$ and the success probability. To do so, we consider the product of the enhancement $\mathcal{D}_\text{NG}$ and the success probability defined as $\mathcal{R} = \mathcal{D}_\text{NG} \times P$. We first plot the product $\mathcal{R}$ as a function of squeezing and transmissivity in Fig.~\ref{prob_contour}. 
	The curves showcasing the highest magnitude values of $\mathcal{R}$ enclose a specific region within the $(r  ,T)$ space, offering the maximum advantage. Experimental implementation can target the squeezing and transmissivity parameters corresponding to this region. Additionally, in Table~\ref{table2}, we present the precise numerical values of the squeezing and transmissivity parameters that maximize the product $\mathcal{R}$.

	\begin{table}[h!]
		\centering
		\caption{\label{table2}
			Maximum value of the product $\mathcal{R}$ and corresponding optimal parameters for different non-Gaussian operations}
		\renewcommand{\arraystretch}{1.5}
		\begin{tabular}{ |c|c |c|c |c|c|c|}
			\hline \hline
			Operation & $10^4 \times \mathcal{R}_\text{max}$  &  $\lambda_\text{opt}$ & $T_\text{opt}$& ${\Delta \phi}_\text{SVS}$ & $\mathcal{D}$ & $10^3 \times P$\\ \hline
			2-PS & 3.6 & 0.38 & 0.55 & 0.23 & 0.05 & 7.9 \\\hline
			2-PC & 5.9 & 0.22 & 0.13 &0.32 &0.12 & 5.0\\ \hline \hline 
		\end{tabular}
	\end{table}
	For comparison, Ref.~\cite{Multistep} considered a squeezing value of $\lambda=0.27$ (2.4 dB) and transmissivity $T=0.9$. The corresponding enhancement in quadrature squeezing turns out to be $\mathcal{D} =0.12$, alongside a   success probability is $P= 2.4\times 10^{-4}$.

	\section{Conclusion}\label{sec:con}
	We examined a   practical scheme for executing three distinct non-Gaussian operations namely PS, PA and PC operations on SVS to distill squeezing. Our findings reveal that both PS and PC operations are effective in squeezing enhancement. Particularly noteworthy is the substantial increase in the quadrature squeezing facilitated by 2-PC operation  within the low squeezing domain of the SVS state. Our practical scheme enables the consideration of success probability, allowing us to furnish optimal parameters for more efficient resource utilization when factoring in this probability. Our study decisively addresses the potential utilization of other non-Gaussian operations beyond the 2-PS operation utilized in Ref.~\cite{Multistep}.

	Furthermore, we have furnished the Wigner characteristic function of the NGSVSs, which holds promise for investigating various other nonclassicality measures~\cite{THAPLIYAL20173178,Priya}. These measures encompass evaluations such as of the Mandel-$Q$ parameter~\cite{Mandel:79,Biswas}, manifestations of antibunching effects~\cite{antibunching}, and assessments of non-Gaussianity~\cite{non-G}. Exploring the impact of these non-Gaussian operations on squeezing distillation in lossy environments   shall be taken elsewhere~\cite{ pra2010,Ulanov2015,Meena2023}.


\begin{thebibliography}{65}%
		\makeatletter
		\providecommand \@ifxundefined [1]{%
			\@ifx{#1\undefined}
		}%
		\providecommand \@ifnum [1]{%
			\ifnum #1\expandafter \@firstoftwo
			\else \expandafter \@secondoftwo
			\fi
		}%
		\providecommand \@ifx [1]{%
			\ifx #1\expandafter \@firstoftwo
			\else \expandafter \@secondoftwo
			\fi
		}%
		\providecommand \natexlab [1]{#1}%
		\providecommand \enquote  [1]{``#1''}%
		\providecommand \bibnamefont  [1]{#1}%
		\providecommand \bibfnamefont [1]{#1}%
		\providecommand \citenamefont [1]{#1}%
		\providecommand \href@noop [0]{\@secondoftwo}%
		\providecommand \href [0]{\begingroup \@sanitize@url \@href}%
		\providecommand \@href[1]{\@@startlink{#1}\@@href}%
		\providecommand \@@href[1]{\endgroup#1\@@endlink}%
		\providecommand \@sanitize@url [0]{\catcode `\\12\catcode `\$12\catcode
			`\&12\catcode `\#12\catcode `\^12\catcode `\_12\catcode `\%12\relax}%
		\providecommand \@@startlink[1]{}%
		\providecommand \@@endlink[0]{}%
		\providecommand \url  [0]{\begingroup\@sanitize@url \@url }%
		\providecommand \@url [1]{\endgroup\@href {#1}{\urlprefix }}%
		\providecommand \urlprefix  [0]{URL }%
		\providecommand \Eprint [0]{\href }%
		\providecommand \doibase [0]{https://doi.org/}%
		\providecommand \selectlanguage [0]{\@gobble}%
		\providecommand \bibinfo  [0]{\@secondoftwo}%
		\providecommand \bibfield  [0]{\@secondoftwo}%
		\providecommand \translation [1]{[#1]}%
		\providecommand \BibitemOpen [0]{}%
		\providecommand \bibitemStop [0]{}%
		\providecommand \bibitemNoStop [0]{.\EOS\space}%
		\providecommand \EOS [0]{\spacefactor3000\relax}%
		\providecommand \BibitemShut  [1]{\csname bibitem#1\endcsname}%
		\let\auto@bib@innerbib\@empty
		\bibitem [{\citenamefont {Braunstein}\ and\ \citenamefont {van
				Loock}(2005)}]{Braunstein}%
		\BibitemOpen
		\bibfield  {author} {\bibinfo {author} {\bibfnamefont {S.~L.}\ \bibnamefont
				{Braunstein}}\ and\ \bibinfo {author} {\bibfnamefont {P.}~\bibnamefont {van
					Loock}},\ }\bibfield  {title} {\bibinfo {title} {Quantum information with
				continuous variables},\ }\href {https://doi.org/10.1103/RevModPhys.77.513}
		{\bibfield  {journal} {\bibinfo  {journal} {Rev. Mod. Phys.}\ }\textbf
			{\bibinfo {volume} {77}},\ \bibinfo {pages} {513} (\bibinfo {year}
			{2005})}\BibitemShut {NoStop}%
		\bibitem [{\citenamefont {Andersen}\ \emph {et~al.}(2010)\citenamefont
			{Andersen}, \citenamefont {Leuchs},\ and\ \citenamefont
			{Silberhorn}}]{Silberhorn}%
		\BibitemOpen
		\bibfield  {author} {\bibinfo {author} {\bibfnamefont {U.}~\bibnamefont
				{Andersen}}, \bibinfo {author} {\bibfnamefont {G.}~\bibnamefont {Leuchs}},\
			and\ \bibinfo {author} {\bibfnamefont {C.}~\bibnamefont {Silberhorn}},\
		}\bibfield  {title} {\bibinfo {title} {Continuous-variable quantum
				information processing},\ }\href
		{https://doi.org/https://doi.org/10.1002/lpor.200910010} {\bibfield
			{journal} {\bibinfo  {journal} {Laser \& Photonics Reviews}\ }\textbf
			{\bibinfo {volume} {4}},\ \bibinfo {pages} {337} (\bibinfo {year}
			{2010})}\BibitemShut {NoStop}%
		\bibitem [{\citenamefont {Weedbrook}\ \emph {et~al.}(2012)\citenamefont
			{Weedbrook}, \citenamefont {Pirandola}, \citenamefont {Garc\'{\i}a-Patr\'on},
			\citenamefont {Cerf}, \citenamefont {Ralph}, \citenamefont {Shapiro},\ and\
			\citenamefont {Lloyd}}]{weedbrook-rmp-2012}%
		\BibitemOpen
		\bibfield  {author} {\bibinfo {author} {\bibfnamefont {C.}~\bibnamefont
				{Weedbrook}}, \bibinfo {author} {\bibfnamefont {S.}~\bibnamefont
				{Pirandola}}, \bibinfo {author} {\bibfnamefont {R.}~\bibnamefont
				{Garc\'{\i}a-Patr\'on}}, \bibinfo {author} {\bibfnamefont {N.~J.}\
				\bibnamefont {Cerf}}, \bibinfo {author} {\bibfnamefont {T.~C.}\ \bibnamefont
				{Ralph}}, \bibinfo {author} {\bibfnamefont {J.~H.}\ \bibnamefont {Shapiro}},\
			and\ \bibinfo {author} {\bibfnamefont {S.}~\bibnamefont {Lloyd}},\ }\bibfield
		{title} {\bibinfo {title} {Gaussian quantum information},\ }\href
		{https://doi.org/10.1103/RevModPhys.84.621} {\bibfield  {journal} {\bibinfo
				{journal} {Rev. Mod. Phys.}\ }\textbf {\bibinfo {volume} {84}},\ \bibinfo
			{pages} {621} (\bibinfo {year} {2012})}\BibitemShut {NoStop}%
		\bibitem [{\citenamefont {Adesso}\ \emph {et~al.}(2014)\citenamefont {Adesso},
			\citenamefont {Ragy},\ and\ \citenamefont {Lee}}]{Adesso}%
		\BibitemOpen
		\bibfield  {author} {\bibinfo {author} {\bibfnamefont {G.}~\bibnamefont
				{Adesso}}, \bibinfo {author} {\bibfnamefont {S.}~\bibnamefont {Ragy}},\ and\
			\bibinfo {author} {\bibfnamefont {A.~R.}\ \bibnamefont {Lee}},\ }\bibfield
		{title} {\bibinfo {title} {Continuous variable quantum information: Gaussian
				states and beyond},\ }\href {https://doi.org/10.1142/S1230161214400010}
		{\bibfield  {journal} {\bibinfo  {journal} {Open Systems \& Information
					Dynamics}\ }\textbf {\bibinfo {volume} {21}},\ \bibinfo {pages} {1440001}
			(\bibinfo {year} {2014})}\BibitemShut {NoStop}%
		\bibitem [{\citenamefont {Furusawa}\ \emph {et~al.}(1998)\citenamefont
			{Furusawa}, \citenamefont {Sørensen}, \citenamefont {Braunstein},
			\citenamefont {Fuchs}, \citenamefont {Kimble},\ and\ \citenamefont
			{Polzik}}]{Furusawa}%
		\BibitemOpen
		\bibfield  {author} {\bibinfo {author} {\bibfnamefont {A.}~\bibnamefont
				{Furusawa}}, \bibinfo {author} {\bibfnamefont {J.~L.}\ \bibnamefont
				{Sørensen}}, \bibinfo {author} {\bibfnamefont {S.~L.}\ \bibnamefont
				{Braunstein}}, \bibinfo {author} {\bibfnamefont {C.~A.}\ \bibnamefont
				{Fuchs}}, \bibinfo {author} {\bibfnamefont {H.~J.}\ \bibnamefont {Kimble}},\
			and\ \bibinfo {author} {\bibfnamefont {E.~S.}\ \bibnamefont {Polzik}},\
		}\bibfield  {title} {\bibinfo {title} {Unconditional quantum teleportation},\
		}\href {https://doi.org/10.1126/science.282.5389.706} {\bibfield  {journal}
			{\bibinfo  {journal} {Science}\ }\textbf {\bibinfo {volume} {282}},\ \bibinfo
			{pages} {706} (\bibinfo {year} {1998})}\BibitemShut {NoStop}%
		\bibitem [{\citenamefont {Bowen}\ \emph {et~al.}(2003)\citenamefont {Bowen},
			\citenamefont {Treps}, \citenamefont {Buchler}, \citenamefont {Schnabel},
			\citenamefont {Ralph}, \citenamefont {Bachor}, \citenamefont {Symul},\ and\
			\citenamefont {Lam}}]{tele}%
		\BibitemOpen
		\bibfield  {author} {\bibinfo {author} {\bibfnamefont {W.~P.}\ \bibnamefont
				{Bowen}}, \bibinfo {author} {\bibfnamefont {N.}~\bibnamefont {Treps}},
			\bibinfo {author} {\bibfnamefont {B.~C.}\ \bibnamefont {Buchler}}, \bibinfo
			{author} {\bibfnamefont {R.}~\bibnamefont {Schnabel}}, \bibinfo {author}
			{\bibfnamefont {T.~C.}\ \bibnamefont {Ralph}}, \bibinfo {author}
			{\bibfnamefont {H.-A.}\ \bibnamefont {Bachor}}, \bibinfo {author}
			{\bibfnamefont {T.}~\bibnamefont {Symul}},\ and\ \bibinfo {author}
			{\bibfnamefont {P.~K.}\ \bibnamefont {Lam}},\ }\bibfield  {title} {\bibinfo
			{title} {Experimental investigation of continuous-variable quantum
				teleportation},\ }\href {https://doi.org/10.1103/PhysRevA.67.032302}
		{\bibfield  {journal} {\bibinfo  {journal} {Phys. Rev. A}\ }\textbf {\bibinfo
				{volume} {67}},\ \bibinfo {pages} {032302} (\bibinfo {year}
			{2003})}\BibitemShut {NoStop}%
		\bibitem [{\citenamefont {Caves}(1981)}]{metro}%
		\BibitemOpen
		\bibfield  {author} {\bibinfo {author} {\bibfnamefont {C.~M.}\ \bibnamefont
				{Caves}},\ }\bibfield  {title} {\bibinfo {title} {Quantum-mechanical noise in
				an interferometer},\ }\href {https://doi.org/10.1103/PhysRevD.23.1693}
		{\bibfield  {journal} {\bibinfo  {journal} {Phys. Rev. D}\ }\textbf {\bibinfo
				{volume} {23}},\ \bibinfo {pages} {1693} (\bibinfo {year}
			{1981})}\BibitemShut {NoStop}%
		\bibitem [{\citenamefont {Giovannetti}\ \emph {et~al.}(2004)\citenamefont
			{Giovannetti}, \citenamefont {Lloyd},\ and\ \citenamefont
			{Maccone}}]{Giovannetti}%
		\BibitemOpen
		\bibfield  {author} {\bibinfo {author} {\bibfnamefont {V.}~\bibnamefont
				{Giovannetti}}, \bibinfo {author} {\bibfnamefont {S.}~\bibnamefont {Lloyd}},\
			and\ \bibinfo {author} {\bibfnamefont {L.}~\bibnamefont {Maccone}},\
		}\bibfield  {title} {\bibinfo {title} {Quantum-enhanced measurements: Beating
				the standard quantum limit},\ }\href
		{https://doi.org/10.1126/science.1104149} {\bibfield  {journal} {\bibinfo
				{journal} {Science}\ }\textbf {\bibinfo {volume} {306}},\ \bibinfo {pages}
			{1330} (\bibinfo {year} {2004})}\BibitemShut {NoStop}%
		\bibitem [{\citenamefont {Gehring}\ \emph {et~al.}(2015)\citenamefont
			{Gehring}, \citenamefont {H{\"a}ndchen}, \citenamefont {Duhme}, \citenamefont
			{Furrer}, \citenamefont {Franz}, \citenamefont {Pacher}, \citenamefont
			{Werner},\ and\ \citenamefont {Schnabel}}]{qkd}%
		\BibitemOpen
		\bibfield  {author} {\bibinfo {author} {\bibfnamefont {T.}~\bibnamefont
				{Gehring}}, \bibinfo {author} {\bibfnamefont {V.}~\bibnamefont
				{H{\"a}ndchen}}, \bibinfo {author} {\bibfnamefont {J.}~\bibnamefont {Duhme}},
			\bibinfo {author} {\bibfnamefont {F.}~\bibnamefont {Furrer}}, \bibinfo
			{author} {\bibfnamefont {T.}~\bibnamefont {Franz}}, \bibinfo {author}
			{\bibfnamefont {C.}~\bibnamefont {Pacher}}, \bibinfo {author} {\bibfnamefont
				{R.~F.}\ \bibnamefont {Werner}},\ and\ \bibinfo {author} {\bibfnamefont
				{R.}~\bibnamefont {Schnabel}},\ }\bibfield  {title} {\bibinfo {title}
			{Implementation of continuous-variable quantum key distribution with
				composable and one-sided-device-independent security against coherent
				attacks},\ }\href {https://doi.org/10.1038/ncomms9795} {\bibfield  {journal}
			{\bibinfo  {journal} {Nature Communications}\ }\textbf {\bibinfo {volume}
				{6}},\ \bibinfo {pages} {8795} (\bibinfo {year} {2015})}\BibitemShut
		{NoStop}%
		\bibitem [{\citenamefont {Gottesman}\ and\ \citenamefont
			{Preskill}(2001)}]{qkd1}%
		\BibitemOpen
		\bibfield  {author} {\bibinfo {author} {\bibfnamefont {D.}~\bibnamefont
				{Gottesman}}\ and\ \bibinfo {author} {\bibfnamefont {J.}~\bibnamefont
				{Preskill}},\ }\bibfield  {title} {\bibinfo {title} {Secure quantum key
				distribution using squeezed states},\ }\href
		{https://doi.org/10.1103/PhysRevA.63.022309} {\bibfield  {journal} {\bibinfo
				{journal} {Phys. Rev. A}\ }\textbf {\bibinfo {volume} {63}},\ \bibinfo
			{pages} {022309} (\bibinfo {year} {2001})}\BibitemShut {NoStop}%
		\bibitem [{\citenamefont {Pirandola}\ \emph {et~al.}(2020)\citenamefont
			{Pirandola}, \citenamefont {Andersen}, \citenamefont {Banchi}, \citenamefont
			{Berta}, \citenamefont {Bunandar}, \citenamefont {Colbeck}, \citenamefont
			{Englund}, \citenamefont {Gehring}, \citenamefont {Lupo}, \citenamefont
			{Ottaviani}, \citenamefont {Pereira}, \citenamefont {Razavi}, \citenamefont
			{Shaari}, \citenamefont {Tomamichel}, \citenamefont {Usenko}, \citenamefont
			{Vallone}, \citenamefont {Villoresi},\ and\ \citenamefont
			{Wallden}}]{Pirandola:20}%
		\BibitemOpen
		\bibfield  {author} {\bibinfo {author} {\bibfnamefont {S.}~\bibnamefont
				{Pirandola}}, \bibinfo {author} {\bibfnamefont {U.~L.}\ \bibnamefont
				{Andersen}}, \bibinfo {author} {\bibfnamefont {L.}~\bibnamefont {Banchi}},
			\bibinfo {author} {\bibfnamefont {M.}~\bibnamefont {Berta}}, \bibinfo
			{author} {\bibfnamefont {D.}~\bibnamefont {Bunandar}}, \bibinfo {author}
			{\bibfnamefont {R.}~\bibnamefont {Colbeck}}, \bibinfo {author} {\bibfnamefont
				{D.}~\bibnamefont {Englund}}, \bibinfo {author} {\bibfnamefont
				{T.}~\bibnamefont {Gehring}}, \bibinfo {author} {\bibfnamefont
				{C.}~\bibnamefont {Lupo}}, \bibinfo {author} {\bibfnamefont {C.}~\bibnamefont
				{Ottaviani}}, \bibinfo {author} {\bibfnamefont {J.~L.}\ \bibnamefont
				{Pereira}}, \bibinfo {author} {\bibfnamefont {M.}~\bibnamefont {Razavi}},
			\bibinfo {author} {\bibfnamefont {J.~S.}\ \bibnamefont {Shaari}}, \bibinfo
			{author} {\bibfnamefont {M.}~\bibnamefont {Tomamichel}}, \bibinfo {author}
			{\bibfnamefont {V.~C.}\ \bibnamefont {Usenko}}, \bibinfo {author}
			{\bibfnamefont {G.}~\bibnamefont {Vallone}}, \bibinfo {author} {\bibfnamefont
				{P.}~\bibnamefont {Villoresi}},\ and\ \bibinfo {author} {\bibfnamefont
				{P.}~\bibnamefont {Wallden}},\ }\bibfield  {title} {\bibinfo {title}
			{Advances in quantum cryptography},\ }\href
		{https://doi.org/10.1364/AOP.361502} {\bibfield  {journal} {\bibinfo
				{journal} {Adv. Opt. Photon.}\ }\textbf {\bibinfo {volume} {12}},\ \bibinfo
			{pages} {1012} (\bibinfo {year} {2020})}\BibitemShut {NoStop}%
		\bibitem [{\citenamefont {Vahlbruch}\ \emph {et~al.}(2008)\citenamefont
			{Vahlbruch}, \citenamefont {Mehmet}, \citenamefont {Chelkowski},
			\citenamefont {Hage}, \citenamefont {Franzen}, \citenamefont {Lastzka},
			\citenamefont {Go\ss{}ler}, \citenamefont {Danzmann},\ and\ \citenamefont
			{Schnabel}}]{Vahlbruch}%
		\BibitemOpen
		\bibfield  {author} {\bibinfo {author} {\bibfnamefont {H.}~\bibnamefont
				{Vahlbruch}}, \bibinfo {author} {\bibfnamefont {M.}~\bibnamefont {Mehmet}},
			\bibinfo {author} {\bibfnamefont {S.}~\bibnamefont {Chelkowski}}, \bibinfo
			{author} {\bibfnamefont {B.}~\bibnamefont {Hage}}, \bibinfo {author}
			{\bibfnamefont {A.}~\bibnamefont {Franzen}}, \bibinfo {author} {\bibfnamefont
				{N.}~\bibnamefont {Lastzka}}, \bibinfo {author} {\bibfnamefont
				{S.}~\bibnamefont {Go\ss{}ler}}, \bibinfo {author} {\bibfnamefont
				{K.}~\bibnamefont {Danzmann}},\ and\ \bibinfo {author} {\bibfnamefont
				{R.}~\bibnamefont {Schnabel}},\ }\bibfield  {title} {\bibinfo {title}
			{Observation of squeezed light with 10-db quantum-noise reduction},\ }\href
		{https://doi.org/10.1103/PhysRevLett.100.033602} {\bibfield  {journal}
			{\bibinfo  {journal} {Phys. Rev. Lett.}\ }\textbf {\bibinfo {volume} {100}},\
			\bibinfo {pages} {033602} (\bibinfo {year} {2008})}\BibitemShut {NoStop}%
		\bibitem [{\citenamefont {Mehmet}\ \emph {et~al.}(2011)\citenamefont {Mehmet},
			\citenamefont {Ast}, \citenamefont {Eberle}, \citenamefont {Steinlechner},
			\citenamefont {Vahlbruch},\ and\ \citenamefont {Schnabel}}]{Mehmet:11}%
		\BibitemOpen
		\bibfield  {author} {\bibinfo {author} {\bibfnamefont {M.}~\bibnamefont
				{Mehmet}}, \bibinfo {author} {\bibfnamefont {S.}~\bibnamefont {Ast}},
			\bibinfo {author} {\bibfnamefont {T.}~\bibnamefont {Eberle}}, \bibinfo
			{author} {\bibfnamefont {S.}~\bibnamefont {Steinlechner}}, \bibinfo {author}
			{\bibfnamefont {H.}~\bibnamefont {Vahlbruch}},\ and\ \bibinfo {author}
			{\bibfnamefont {R.}~\bibnamefont {Schnabel}},\ }\bibfield  {title} {\bibinfo
			{title} {Squeezed light at 1550 nm with a quantum noise reduction of 12.3
				db},\ }\href {https://doi.org/10.1364/OE.19.025763} {\bibfield  {journal}
			{\bibinfo  {journal} {Opt. Express}\ }\textbf {\bibinfo {volume} {19}},\
			\bibinfo {pages} {25763} (\bibinfo {year} {2011})}\BibitemShut {NoStop}%
		\bibitem [{\citenamefont {Vahlbruch}\ \emph {et~al.}(2016)\citenamefont
			{Vahlbruch}, \citenamefont {Mehmet}, \citenamefont {Danzmann},\ and\
			\citenamefont {Schnabel}}]{15db}%
		\BibitemOpen
		\bibfield  {author} {\bibinfo {author} {\bibfnamefont {H.}~\bibnamefont
				{Vahlbruch}}, \bibinfo {author} {\bibfnamefont {M.}~\bibnamefont {Mehmet}},
			\bibinfo {author} {\bibfnamefont {K.}~\bibnamefont {Danzmann}},\ and\
			\bibinfo {author} {\bibfnamefont {R.}~\bibnamefont {Schnabel}},\ }\bibfield
		{title} {\bibinfo {title} {Detection of 15 db squeezed states of light and
				their application for the absolute calibration of photoelectric quantum
				efficiency},\ }\href {https://doi.org/10.1103/PhysRevLett.117.110801}
		{\bibfield  {journal} {\bibinfo  {journal} {Phys. Rev. Lett.}\ }\textbf
			{\bibinfo {volume} {117}},\ \bibinfo {pages} {110801} (\bibinfo {year}
			{2016})}\BibitemShut {NoStop}%
		\bibitem [{\citenamefont {Grebien}\ \emph {et~al.}(2022)\citenamefont
			{Grebien}, \citenamefont {G\"ottsch}, \citenamefont {Hage}, \citenamefont
			{Fiur\'a\ifmmode~\check{s}\else \v{s}\fi{}ek},\ and\ \citenamefont
			{Schnabel}}]{Multistep}%
		\BibitemOpen
		\bibfield  {author} {\bibinfo {author} {\bibfnamefont {S.}~\bibnamefont
				{Grebien}}, \bibinfo {author} {\bibfnamefont {J.}~\bibnamefont {G\"ottsch}},
			\bibinfo {author} {\bibfnamefont {B.}~\bibnamefont {Hage}}, \bibinfo {author}
			{\bibfnamefont {J.}~\bibnamefont {Fiur\'a\ifmmode~\check{s}\else
					\v{s}\fi{}ek}},\ and\ \bibinfo {author} {\bibfnamefont {R.}~\bibnamefont
				{Schnabel}},\ }\bibfield  {title} {\bibinfo {title} {Multistep two-copy
				distillation of squeezed states via two-photon subtraction},\ }\href
		{https://doi.org/10.1103/PhysRevLett.129.273604} {\bibfield  {journal}
			{\bibinfo  {journal} {Phys. Rev. Lett.}\ }\textbf {\bibinfo {volume} {129}},\
			\bibinfo {pages} {273604} (\bibinfo {year} {2022})}\BibitemShut {NoStop}%
		\bibitem [{\citenamefont {Magrini}\ \emph {et~al.}(2022)\citenamefont
			{Magrini}, \citenamefont {Camarena-Ch\'avez}, \citenamefont {Bach},
			\citenamefont {Johnson},\ and\ \citenamefont {Aspelmeyer}}]{1.1dB}%
		\BibitemOpen
		\bibfield  {author} {\bibinfo {author} {\bibfnamefont {L.}~\bibnamefont
				{Magrini}}, \bibinfo {author} {\bibfnamefont {V.~A.}\ \bibnamefont
				{Camarena-Ch\'avez}}, \bibinfo {author} {\bibfnamefont {C.}~\bibnamefont
				{Bach}}, \bibinfo {author} {\bibfnamefont {A.}~\bibnamefont {Johnson}},\ and\
			\bibinfo {author} {\bibfnamefont {M.}~\bibnamefont {Aspelmeyer}},\ }\bibfield
		{title} {\bibinfo {title} {Squeezed light from a levitated nanoparticle at
				room temperature},\ }\href {https://doi.org/10.1103/PhysRevLett.129.053601}
		{\bibfield  {journal} {\bibinfo  {journal} {Phys. Rev. Lett.}\ }\textbf
			{\bibinfo {volume} {129}},\ \bibinfo {pages} {053601} (\bibinfo {year}
			{2022})}\BibitemShut {NoStop}%
		\bibitem [{\citenamefont {Militaru}\ \emph {et~al.}(2022)\citenamefont
			{Militaru}, \citenamefont {Rossi}, \citenamefont {Tebbenjohanns},
			\citenamefont {Romero-Isart}, \citenamefont {Frimmer},\ and\ \citenamefont
			{Novotny}}]{1.3dB}%
		\BibitemOpen
		\bibfield  {author} {\bibinfo {author} {\bibfnamefont {A.}~\bibnamefont
				{Militaru}}, \bibinfo {author} {\bibfnamefont {M.}~\bibnamefont {Rossi}},
			\bibinfo {author} {\bibfnamefont {F.}~\bibnamefont {Tebbenjohanns}}, \bibinfo
			{author} {\bibfnamefont {O.}~\bibnamefont {Romero-Isart}}, \bibinfo {author}
			{\bibfnamefont {M.}~\bibnamefont {Frimmer}},\ and\ \bibinfo {author}
			{\bibfnamefont {L.}~\bibnamefont {Novotny}},\ }\bibfield  {title} {\bibinfo
			{title} {Ponderomotive squeezing of light by a levitated nanoparticle in free
				space},\ }\href {https://doi.org/10.1103/PhysRevLett.129.053602} {\bibfield
			{journal} {\bibinfo  {journal} {Phys. Rev. Lett.}\ }\textbf {\bibinfo
				{volume} {129}},\ \bibinfo {pages} {053602} (\bibinfo {year}
			{2022})}\BibitemShut {NoStop}%
		\bibitem [{\citenamefont {Kraus}\ \emph {et~al.}(2003)\citenamefont {Kraus},
			\citenamefont {Hammerer}, \citenamefont {Giedke},\ and\ \citenamefont
			{Cirac}}]{Kraus}%
		\BibitemOpen
		\bibfield  {author} {\bibinfo {author} {\bibfnamefont {B.}~\bibnamefont
				{Kraus}}, \bibinfo {author} {\bibfnamefont {K.}~\bibnamefont {Hammerer}},
			\bibinfo {author} {\bibfnamefont {G.}~\bibnamefont {Giedke}},\ and\ \bibinfo
			{author} {\bibfnamefont {J.~I.}\ \bibnamefont {Cirac}},\ }\bibfield  {title}
		{\bibinfo {title} {Entanglement generation and hamiltonian simulation in
				continuous-variable systems},\ }\href
		{https://doi.org/10.1103/PhysRevA.67.042314} {\bibfield  {journal} {\bibinfo
				{journal} {Phys. Rev. A}\ }\textbf {\bibinfo {volume} {67}},\ \bibinfo
			{pages} {042314} (\bibinfo {year} {2003})}\BibitemShut {NoStop}%
		\bibitem [{\citenamefont {Eisert}\ \emph {et~al.}(2002)\citenamefont {Eisert},
			\citenamefont {Scheel},\ and\ \citenamefont {Plenio}}]{Eisert}%
		\BibitemOpen
		\bibfield  {author} {\bibinfo {author} {\bibfnamefont {J.}~\bibnamefont
				{Eisert}}, \bibinfo {author} {\bibfnamefont {S.}~\bibnamefont {Scheel}},\
			and\ \bibinfo {author} {\bibfnamefont {M.~B.}\ \bibnamefont {Plenio}},\
		}\bibfield  {title} {\bibinfo {title} {Distilling gaussian states with
				gaussian operations is impossible},\ }\href
		{https://doi.org/10.1103/PhysRevLett.89.137903} {\bibfield  {journal}
			{\bibinfo  {journal} {Phys. Rev. Lett.}\ }\textbf {\bibinfo {volume} {89}},\
			\bibinfo {pages} {137903} (\bibinfo {year} {2002})}\BibitemShut {NoStop}%
		\bibitem [{\citenamefont {Fiur\'a\ifmmode~\check{s}\else
				\v{s}\fi{}ek}(2002)}]{prl2002}%
		\BibitemOpen
		\bibfield  {author} {\bibinfo {author} {\bibfnamefont {J.}~\bibnamefont
				{Fiur\'a\ifmmode~\check{s}\else \v{s}\fi{}ek}},\ }\bibfield  {title}
		{\bibinfo {title} {Gaussian transformations and distillation of entangled
				gaussian states},\ }\href {https://doi.org/10.1103/PhysRevLett.89.137904}
		{\bibfield  {journal} {\bibinfo  {journal} {Phys. Rev. Lett.}\ }\textbf
			{\bibinfo {volume} {89}},\ \bibinfo {pages} {137904} (\bibinfo {year}
			{2002})}\BibitemShut {NoStop}%
		\bibitem [{\citenamefont {Giedke}\ and\ \citenamefont
			{Ignacio~Cirac}(2002)}]{Giedke}%
		\BibitemOpen
		\bibfield  {author} {\bibinfo {author} {\bibfnamefont {G.}~\bibnamefont
				{Giedke}}\ and\ \bibinfo {author} {\bibfnamefont {J.}~\bibnamefont
				{Ignacio~Cirac}},\ }\bibfield  {title} {\bibinfo {title} {Characterization of
				gaussian operations and distillation of gaussian states},\ }\href
		{https://doi.org/10.1103/PhysRevA.66.032316} {\bibfield  {journal} {\bibinfo
				{journal} {Phys. Rev. A}\ }\textbf {\bibinfo {volume} {66}},\ \bibinfo
			{pages} {032316} (\bibinfo {year} {2002})}\BibitemShut {NoStop}%
		\bibitem [{\citenamefont {Ourjoumtsev}\ \emph {et~al.}(2007)\citenamefont
			{Ourjoumtsev}, \citenamefont {Dantan}, \citenamefont {Tualle-Brouri},\ and\
			\citenamefont {Grangier}}]{Ourjoumtsev}%
		\BibitemOpen
		\bibfield  {author} {\bibinfo {author} {\bibfnamefont {A.}~\bibnamefont
				{Ourjoumtsev}}, \bibinfo {author} {\bibfnamefont {A.}~\bibnamefont {Dantan}},
			\bibinfo {author} {\bibfnamefont {R.}~\bibnamefont {Tualle-Brouri}},\ and\
			\bibinfo {author} {\bibfnamefont {P.}~\bibnamefont {Grangier}},\ }\bibfield
		{title} {\bibinfo {title} {Increasing entanglement between gaussian states by
				coherent photon subtraction},\ }\href
		{https://doi.org/10.1103/PhysRevLett.98.030502} {\bibfield  {journal}
			{\bibinfo  {journal} {Phys. Rev. Lett.}\ }\textbf {\bibinfo {volume} {98}},\
			\bibinfo {pages} {030502} (\bibinfo {year} {2007})}\BibitemShut {NoStop}%
		\bibitem [{\citenamefont {Takahashi}\ \emph {et~al.}(2010)\citenamefont
			{Takahashi}, \citenamefont {Neergaard-Nielsen}, \citenamefont {Takeuchi},
			\citenamefont {Takeoka}, \citenamefont {Hayasaka}, \citenamefont {Furusawa},\
			and\ \citenamefont {Sasaki}}]{Takahashi2010}%
		\BibitemOpen
		\bibfield  {author} {\bibinfo {author} {\bibfnamefont {H.}~\bibnamefont
				{Takahashi}}, \bibinfo {author} {\bibfnamefont {J.~S.}\ \bibnamefont
				{Neergaard-Nielsen}}, \bibinfo {author} {\bibfnamefont {M.}~\bibnamefont
				{Takeuchi}}, \bibinfo {author} {\bibfnamefont {M.}~\bibnamefont {Takeoka}},
			\bibinfo {author} {\bibfnamefont {K.}~\bibnamefont {Hayasaka}}, \bibinfo
			{author} {\bibfnamefont {A.}~\bibnamefont {Furusawa}},\ and\ \bibinfo
			{author} {\bibfnamefont {M.}~\bibnamefont {Sasaki}},\ }\bibfield  {title}
		{\bibinfo {title} {Entanglement distillation from gaussian input states},\
		}\href {https://doi.org/10.1038/nphoton.2010.1} {\bibfield  {journal}
			{\bibinfo  {journal} {Nature Photonics}\ }\textbf {\bibinfo {volume} {4}},\
			\bibinfo {pages} {178} (\bibinfo {year} {2010})}\BibitemShut {NoStop}%
		\bibitem [{\citenamefont {Kurochkin}\ \emph {et~al.}(2014)\citenamefont
			{Kurochkin}, \citenamefont {Prasad},\ and\ \citenamefont
			{Lvovsky}}]{Lvovsky}%
		\BibitemOpen
		\bibfield  {author} {\bibinfo {author} {\bibfnamefont {Y.}~\bibnamefont
				{Kurochkin}}, \bibinfo {author} {\bibfnamefont {A.~S.}\ \bibnamefont
				{Prasad}},\ and\ \bibinfo {author} {\bibfnamefont {A.~I.}\ \bibnamefont
				{Lvovsky}},\ }\bibfield  {title} {\bibinfo {title} {Distillation of the
				two-mode squeezed state},\ }\href
		{https://doi.org/10.1103/PhysRevLett.112.070402} {\bibfield  {journal}
			{\bibinfo  {journal} {Phys. Rev. Lett.}\ }\textbf {\bibinfo {volume} {112}},\
			\bibinfo {pages} {070402} (\bibinfo {year} {2014})}\BibitemShut {NoStop}%
		\bibitem [{\citenamefont {Dirmeier}\ \emph {et~al.}(2020)\citenamefont
			{Dirmeier}, \citenamefont {Tiedau}, \citenamefont {Khan}, \citenamefont
			{Ansari}, \citenamefont {M\"{u}ller}, \citenamefont {Silberhorn},
			\citenamefont {Marquardt},\ and\ \citenamefont {Leuchs}}]{Dirmeier:20}%
		\BibitemOpen
		\bibfield  {author} {\bibinfo {author} {\bibfnamefont {T.}~\bibnamefont
				{Dirmeier}}, \bibinfo {author} {\bibfnamefont {J.}~\bibnamefont {Tiedau}},
			\bibinfo {author} {\bibfnamefont {I.}~\bibnamefont {Khan}}, \bibinfo {author}
			{\bibfnamefont {V.}~\bibnamefont {Ansari}}, \bibinfo {author} {\bibfnamefont
				{C.~R.}\ \bibnamefont {M\"{u}ller}}, \bibinfo {author} {\bibfnamefont
				{C.}~\bibnamefont {Silberhorn}}, \bibinfo {author} {\bibfnamefont
				{C.}~\bibnamefont {Marquardt}},\ and\ \bibinfo {author} {\bibfnamefont
				{G.}~\bibnamefont {Leuchs}},\ }\bibfield  {title} {\bibinfo {title}
			{Distillation of squeezing using an engineered pulsed parametric
				down-conversion source},\ }\href {https://doi.org/10.1364/OE.402178}
		{\bibfield  {journal} {\bibinfo  {journal} {Opt. Express}\ }\textbf {\bibinfo
				{volume} {28}},\ \bibinfo {pages} {30784} (\bibinfo {year}
			{2020})}\BibitemShut {NoStop}%
		\bibitem [{\citenamefont {Birrittella}\ \emph {et~al.}(2012)\citenamefont
			{Birrittella}, \citenamefont {Mimih},\ and\ \citenamefont
			{Gerry}}]{gerryc-pra-2012}%
		\BibitemOpen
		\bibfield  {author} {\bibinfo {author} {\bibfnamefont {R.}~\bibnamefont
				{Birrittella}}, \bibinfo {author} {\bibfnamefont {J.}~\bibnamefont {Mimih}},\
			and\ \bibinfo {author} {\bibfnamefont {C.~C.}\ \bibnamefont {Gerry}},\
		}\bibfield  {title} {\bibinfo {title} {Multiphoton quantum interference at a
				beam splitter and the approach to heisenberg-limited interferometry},\ }\href
		{https://doi.org/10.1103/PhysRevA.86.063828} {\bibfield  {journal} {\bibinfo
				{journal} {Phys. Rev. A}\ }\textbf {\bibinfo {volume} {86}},\ \bibinfo
			{pages} {063828} (\bibinfo {year} {2012})}\BibitemShut {NoStop}%
		\bibitem [{\citenamefont {Carranza}\ and\ \citenamefont
			{Gerry}(2012)}]{josab-2012}%
		\BibitemOpen
		\bibfield  {author} {\bibinfo {author} {\bibfnamefont {R.}~\bibnamefont
				{Carranza}}\ and\ \bibinfo {author} {\bibfnamefont {C.~C.}\ \bibnamefont
				{Gerry}},\ }\bibfield  {title} {\bibinfo {title} {Photon-subtracted two-mode
				squeezed vacuum states and applications to quantum optical interferometry},\
		}\href {https://doi.org/10.1364/JOSAB.29.002581} {\bibfield  {journal}
			{\bibinfo  {journal} {J. Opt. Soc. Am. B}\ }\textbf {\bibinfo {volume}
				{29}},\ \bibinfo {pages} {2581} (\bibinfo {year} {2012})}\BibitemShut
		{NoStop}%
		\bibitem [{\citenamefont {Braun}\ \emph {et~al.}(2014)\citenamefont {Braun},
			\citenamefont {Jian}, \citenamefont {Pinel},\ and\ \citenamefont
			{Treps}}]{braun-pra-2014}%
		\BibitemOpen
		\bibfield  {author} {\bibinfo {author} {\bibfnamefont {D.}~\bibnamefont
				{Braun}}, \bibinfo {author} {\bibfnamefont {P.}~\bibnamefont {Jian}},
			\bibinfo {author} {\bibfnamefont {O.}~\bibnamefont {Pinel}},\ and\ \bibinfo
			{author} {\bibfnamefont {N.}~\bibnamefont {Treps}},\ }\bibfield  {title}
		{\bibinfo {title} {Precision measurements with photon-subtracted or
				photon-added gaussian states},\ }\href
		{https://doi.org/10.1103/PhysRevA.90.013821} {\bibfield  {journal} {\bibinfo
				{journal} {Phys. Rev. A}\ }\textbf {\bibinfo {volume} {90}},\ \bibinfo
			{pages} {013821} (\bibinfo {year} {2014})}\BibitemShut {NoStop}%
		\bibitem [{\citenamefont {Ouyang}\ \emph {et~al.}(2016)\citenamefont {Ouyang},
			\citenamefont {Wang},\ and\ \citenamefont {Zhang}}]{josab-2016}%
		\BibitemOpen
		\bibfield  {author} {\bibinfo {author} {\bibfnamefont {Y.}~\bibnamefont
				{Ouyang}}, \bibinfo {author} {\bibfnamefont {S.}~\bibnamefont {Wang}},\ and\
			\bibinfo {author} {\bibfnamefont {L.}~\bibnamefont {Zhang}},\ }\bibfield
		{title} {\bibinfo {title} {Quantum optical interferometry via the
				photon-added two-mode squeezed vacuum states},\ }\href
		{https://doi.org/10.1364/JOSAB.33.001373} {\bibfield  {journal} {\bibinfo
				{journal} {J. Opt. Soc. Am. B}\ }\textbf {\bibinfo {volume} {33}},\ \bibinfo
			{pages} {1373} (\bibinfo {year} {2016})}\BibitemShut {NoStop}%
		\bibitem [{\citenamefont {Zhang}\ \emph {et~al.}(2021)\citenamefont {Zhang},
			\citenamefont {Ye}, \citenamefont {Wei}, \citenamefont {Xia}, \citenamefont
			{Chang}, \citenamefont {Liao},\ and\ \citenamefont
			{Hu}}]{pra-catalysis-2021}%
		\BibitemOpen
		\bibfield  {author} {\bibinfo {author} {\bibfnamefont {H.}~\bibnamefont
				{Zhang}}, \bibinfo {author} {\bibfnamefont {W.}~\bibnamefont {Ye}}, \bibinfo
			{author} {\bibfnamefont {C.}~\bibnamefont {Wei}}, \bibinfo {author}
			{\bibfnamefont {Y.}~\bibnamefont {Xia}}, \bibinfo {author} {\bibfnamefont
				{S.}~\bibnamefont {Chang}}, \bibinfo {author} {\bibfnamefont
				{Z.}~\bibnamefont {Liao}},\ and\ \bibinfo {author} {\bibfnamefont
				{L.}~\bibnamefont {Hu}},\ }\bibfield  {title} {\bibinfo {title} {Improved
				phase sensitivity in a quantum optical interferometer based on multiphoton
				catalytic two-mode squeezed vacuum states},\ }\href
		{https://doi.org/10.1103/PhysRevA.103.013705} {\bibfield  {journal} {\bibinfo
				{journal} {Phys. Rev. A}\ }\textbf {\bibinfo {volume} {103}},\ \bibinfo
			{pages} {013705} (\bibinfo {year} {2021})}\BibitemShut {NoStop}%
		\bibitem [{\citenamefont {Tan}\ \emph {et~al.}(2008)\citenamefont {Tan},
			\citenamefont {Erkmen}, \citenamefont {Giovannetti}, \citenamefont {Guha},
			\citenamefont {Lloyd}, \citenamefont {Maccone}, \citenamefont {Pirandola},\
			and\ \citenamefont {Shapiro}}]{ill2008}%
		\BibitemOpen
		\bibfield  {author} {\bibinfo {author} {\bibfnamefont {S.-H.}\ \bibnamefont
				{Tan}}, \bibinfo {author} {\bibfnamefont {B.~I.}\ \bibnamefont {Erkmen}},
			\bibinfo {author} {\bibfnamefont {V.}~\bibnamefont {Giovannetti}}, \bibinfo
			{author} {\bibfnamefont {S.}~\bibnamefont {Guha}}, \bibinfo {author}
			{\bibfnamefont {S.}~\bibnamefont {Lloyd}}, \bibinfo {author} {\bibfnamefont
				{L.}~\bibnamefont {Maccone}}, \bibinfo {author} {\bibfnamefont
				{S.}~\bibnamefont {Pirandola}},\ and\ \bibinfo {author} {\bibfnamefont
				{J.~H.}\ \bibnamefont {Shapiro}},\ }\bibfield  {title} {\bibinfo {title}
			{Quantum illumination with gaussian states},\ }\href
		{https://doi.org/10.1103/PhysRevLett.101.253601} {\bibfield  {journal}
			{\bibinfo  {journal} {Phys. Rev. Lett.}\ }\textbf {\bibinfo {volume} {101}},\
			\bibinfo {pages} {253601} (\bibinfo {year} {2008})}\BibitemShut {NoStop}%
		\bibitem [{\citenamefont {Lopaeva}\ \emph {et~al.}(2013)\citenamefont
			{Lopaeva}, \citenamefont {Ruo~Berchera}, \citenamefont {Degiovanni},
			\citenamefont {Olivares}, \citenamefont {Brida},\ and\ \citenamefont
			{Genovese}}]{ill2013}%
		\BibitemOpen
		\bibfield  {author} {\bibinfo {author} {\bibfnamefont {E.~D.}\ \bibnamefont
				{Lopaeva}}, \bibinfo {author} {\bibfnamefont {I.}~\bibnamefont
				{Ruo~Berchera}}, \bibinfo {author} {\bibfnamefont {I.~P.}\ \bibnamefont
				{Degiovanni}}, \bibinfo {author} {\bibfnamefont {S.}~\bibnamefont
				{Olivares}}, \bibinfo {author} {\bibfnamefont {G.}~\bibnamefont {Brida}},\
			and\ \bibinfo {author} {\bibfnamefont {M.}~\bibnamefont {Genovese}},\
		}\bibfield  {title} {\bibinfo {title} {Experimental realization of quantum
				illumination},\ }\href {https://doi.org/10.1103/PhysRevLett.110.153603}
		{\bibfield  {journal} {\bibinfo  {journal} {Phys. Rev. Lett.}\ }\textbf
			{\bibinfo {volume} {110}},\ \bibinfo {pages} {153603} (\bibinfo {year}
			{2013})}\BibitemShut {NoStop}%
		\bibitem [{\citenamefont {Kumar}\ \emph
			{et~al.}(2023{\natexlab{a}})\citenamefont {Kumar}, \citenamefont {Rishabh},\
			and\ \citenamefont {Arora}}]{metro-thermal-arxiv}%
		\BibitemOpen
		\bibfield  {author} {\bibinfo {author} {\bibfnamefont {C.}~\bibnamefont
				{Kumar}}, \bibinfo {author} {\bibnamefont {Rishabh}},\ and\ \bibinfo {author}
			{\bibfnamefont {S.}~\bibnamefont {Arora}},\ }\bibfield  {title} {\bibinfo
			{title} {Enhanced phase estimation in parity-detection-based mach–zehnder
				interferometer using non-gaussian two-mode squeezed thermal input state},\
		}\href {https://doi.org/https://doi.org/10.1002/andp.202300117} {\bibfield
			{journal} {\bibinfo  {journal} {Annalen der Physik}\ }\textbf {\bibinfo
				{volume} {535}},\ \bibinfo {pages} {2300117} (\bibinfo {year}
			{2023}{\natexlab{a}})}\BibitemShut {NoStop}%
		\bibitem [{\citenamefont {Kumar}\ \emph
			{et~al.}(2023{\natexlab{b}})\citenamefont {Kumar}, \citenamefont {Rishabh},
			\citenamefont {Sharma},\ and\ \citenamefont {Arora}}]{ngsvs-arxiv}%
		\BibitemOpen
		\bibfield  {author} {\bibinfo {author} {\bibfnamefont {C.}~\bibnamefont
				{Kumar}}, \bibinfo {author} {\bibnamefont {Rishabh}}, \bibinfo {author}
			{\bibfnamefont {M.}~\bibnamefont {Sharma}},\ and\ \bibinfo {author}
			{\bibfnamefont {S.}~\bibnamefont {Arora}},\ }\bibfield  {title} {\bibinfo
			{title} {Parity-detection-based mach-zehnder interferometry with coherent and
				non-gaussian squeezed vacuum states as inputs},\ }\href
		{https://doi.org/10.1103/PhysRevA.108.012605} {\bibfield  {journal} {\bibinfo
				{journal} {Phys. Rev. A}\ }\textbf {\bibinfo {volume} {108}},\ \bibinfo
			{pages} {012605} (\bibinfo {year} {2023}{\natexlab{b}})}\BibitemShut
		{NoStop}%
		\bibitem [{\citenamefont {Opatrn\'y}\ \emph {et~al.}(2000)\citenamefont
			{Opatrn\'y}, \citenamefont {Kurizki},\ and\ \citenamefont
			{Welsch}}]{tel2000}%
		\BibitemOpen
		\bibfield  {author} {\bibinfo {author} {\bibfnamefont {T.}~\bibnamefont
				{Opatrn\'y}}, \bibinfo {author} {\bibfnamefont {G.}~\bibnamefont {Kurizki}},\
			and\ \bibinfo {author} {\bibfnamefont {D.-G.}\ \bibnamefont {Welsch}},\
		}\bibfield  {title} {\bibinfo {title} {Improvement on teleportation of
				continuous variables by photon subtraction via conditional measurement},\
		}\href {https://doi.org/10.1103/PhysRevA.61.032302} {\bibfield  {journal}
			{\bibinfo  {journal} {Phys. Rev. A}\ }\textbf {\bibinfo {volume} {61}},\
			\bibinfo {pages} {032302} (\bibinfo {year} {2000})}\BibitemShut {NoStop}%
		\bibitem [{\citenamefont {Dell'Anno}\ \emph {et~al.}(2007)\citenamefont
			{Dell'Anno}, \citenamefont {De~Siena}, \citenamefont {Albano},\ and\
			\citenamefont {Illuminati}}]{dellanno-2007}%
		\BibitemOpen
		\bibfield  {author} {\bibinfo {author} {\bibfnamefont {F.}~\bibnamefont
				{Dell'Anno}}, \bibinfo {author} {\bibfnamefont {S.}~\bibnamefont {De~Siena}},
			\bibinfo {author} {\bibfnamefont {L.}~\bibnamefont {Albano}},\ and\ \bibinfo
			{author} {\bibfnamefont {F.}~\bibnamefont {Illuminati}},\ }\bibfield  {title}
		{\bibinfo {title} {Continuous-variable quantum teleportation with
				non-gaussian resources},\ }\href {https://doi.org/10.1103/PhysRevA.76.022301}
		{\bibfield  {journal} {\bibinfo  {journal} {Phys. Rev. A}\ }\textbf {\bibinfo
				{volume} {76}},\ \bibinfo {pages} {022301} (\bibinfo {year}
			{2007})}\BibitemShut {NoStop}%
		\bibitem [{\citenamefont {Yang}\ and\ \citenamefont {Li}(2009)}]{tel2009}%
		\BibitemOpen
		\bibfield  {author} {\bibinfo {author} {\bibfnamefont {Y.}~\bibnamefont
				{Yang}}\ and\ \bibinfo {author} {\bibfnamefont {F.-L.}\ \bibnamefont {Li}},\
		}\bibfield  {title} {\bibinfo {title} {Entanglement properties of
				non-gaussian resources generated via photon subtraction and addition and
				continuous-variable quantum-teleportation improvement},\ }\href
		{https://doi.org/10.1103/PhysRevA.80.022315} {\bibfield  {journal} {\bibinfo
				{journal} {Phys. Rev. A}\ }\textbf {\bibinfo {volume} {80}},\ \bibinfo
			{pages} {022315} (\bibinfo {year} {2009})}\BibitemShut {NoStop}%
		\bibitem [{\citenamefont {Xu}(2015)}]{catalysis15}%
		\BibitemOpen
		\bibfield  {author} {\bibinfo {author} {\bibfnamefont {X.-x.}\ \bibnamefont
				{Xu}},\ }\bibfield  {title} {\bibinfo {title} {Enhancing quantum entanglement
				and quantum teleportation for two-mode squeezed vacuum state by local
				quantum-optical catalysis},\ }\href
		{https://doi.org/10.1103/PhysRevA.92.012318} {\bibfield  {journal} {\bibinfo
				{journal} {Phys. Rev. A}\ }\textbf {\bibinfo {volume} {92}},\ \bibinfo
			{pages} {012318} (\bibinfo {year} {2015})}\BibitemShut {NoStop}%
		\bibitem [{\citenamefont {Hu}\ \emph {et~al.}(2017)\citenamefont {Hu},
			\citenamefont {Liao},\ and\ \citenamefont {Zubairy}}]{catalysis17}%
		\BibitemOpen
		\bibfield  {author} {\bibinfo {author} {\bibfnamefont {L.}~\bibnamefont
				{Hu}}, \bibinfo {author} {\bibfnamefont {Z.}~\bibnamefont {Liao}},\ and\
			\bibinfo {author} {\bibfnamefont {M.~S.}\ \bibnamefont {Zubairy}},\
		}\bibfield  {title} {\bibinfo {title} {Continuous-variable entanglement via
				multiphoton catalysis},\ }\href {https://doi.org/10.1103/PhysRevA.95.012310}
		{\bibfield  {journal} {\bibinfo  {journal} {Phys. Rev. A}\ }\textbf {\bibinfo
				{volume} {95}},\ \bibinfo {pages} {012310} (\bibinfo {year}
			{2017})}\BibitemShut {NoStop}%
		\bibitem [{\citenamefont {Wang}\ \emph {et~al.}(2015)\citenamefont {Wang},
			\citenamefont {Hou}, \citenamefont {Chen},\ and\ \citenamefont
			{Xu}}]{wang2015}%
		\BibitemOpen
		\bibfield  {author} {\bibinfo {author} {\bibfnamefont {S.}~\bibnamefont
				{Wang}}, \bibinfo {author} {\bibfnamefont {L.-L.}\ \bibnamefont {Hou}},
			\bibinfo {author} {\bibfnamefont {X.-F.}\ \bibnamefont {Chen}},\ and\
			\bibinfo {author} {\bibfnamefont {X.-F.}\ \bibnamefont {Xu}},\ }\bibfield
		{title} {\bibinfo {title} {Continuous-variable quantum teleportation with
				non-gaussian entangled states generated via multiple-photon subtraction and
				addition},\ }\href {https://doi.org/10.1103/PhysRevA.91.063832} {\bibfield
			{journal} {\bibinfo  {journal} {Phys. Rev. A}\ }\textbf {\bibinfo {volume}
				{91}},\ \bibinfo {pages} {063832} (\bibinfo {year} {2015})}\BibitemShut
		{NoStop}%
		\bibitem [{\citenamefont {Kumar}\ and\ \citenamefont
			{Arora}(2023)}]{tele-2023}%
		\BibitemOpen
		\bibfield  {author} {\bibinfo {author} {\bibfnamefont {C.}~\bibnamefont
				{Kumar}}\ and\ \bibinfo {author} {\bibfnamefont {S.}~\bibnamefont {Arora}},\
		}\bibfield  {title} {\bibinfo {title} {Success probability and performance
				optimization in non-gaussian continuous-variable quantum teleportation},\
		}\href {https://doi.org/10.1103/PhysRevA.107.012418} {\bibfield  {journal}
			{\bibinfo  {journal} {Phys. Rev. A}\ }\textbf {\bibinfo {volume} {107}},\
			\bibinfo {pages} {012418} (\bibinfo {year} {2023})}\BibitemShut {NoStop}%
		\bibitem [{\citenamefont {Kumar}\ \emph {et~al.}(2022)\citenamefont {Kumar},
			\citenamefont {Sharma},\ and\ \citenamefont {Arora}}]{noisytele}%
		\BibitemOpen
		\bibfield  {author} {\bibinfo {author} {\bibfnamefont {C.}~\bibnamefont
				{Kumar}}, \bibinfo {author} {\bibfnamefont {M.}~\bibnamefont {Sharma}},\ and\
			\bibinfo {author} {\bibfnamefont {S.}~\bibnamefont {Arora}},\ }\bibfield
		{title} {\bibinfo {title} {Continuous variable quantum teleportation in a
				dissipative environment: Comparison of non-gaussian operations before and
				after noisy channel},\ }\href {https://doi.org/10.48550/arXiv.2212.13133}
		{\bibfield  {journal} {\bibinfo  {journal} {arXiv:2212.13133}\ } (\bibinfo
			{year} {2022})}\BibitemShut {NoStop}%
		\bibitem [{\citenamefont {Vidrighin}\ \emph {et~al.}(2016)\citenamefont
			{Vidrighin}, \citenamefont {Dahlsten}, \citenamefont {Barbieri},
			\citenamefont {Kim}, \citenamefont {Vedral},\ and\ \citenamefont
			{Walmsley}}]{Walmsley}%
		\BibitemOpen
		\bibfield  {author} {\bibinfo {author} {\bibfnamefont {M.~D.}\ \bibnamefont
				{Vidrighin}}, \bibinfo {author} {\bibfnamefont {O.}~\bibnamefont {Dahlsten}},
			\bibinfo {author} {\bibfnamefont {M.}~\bibnamefont {Barbieri}}, \bibinfo
			{author} {\bibfnamefont {M.~S.}\ \bibnamefont {Kim}}, \bibinfo {author}
			{\bibfnamefont {V.}~\bibnamefont {Vedral}},\ and\ \bibinfo {author}
			{\bibfnamefont {I.~A.}\ \bibnamefont {Walmsley}},\ }\bibfield  {title}
		{\bibinfo {title} {Photonic maxwell's demon},\ }\href
		{https://doi.org/10.1103/PhysRevLett.116.050401} {\bibfield  {journal}
			{\bibinfo  {journal} {Phys. Rev. Lett.}\ }\textbf {\bibinfo {volume} {116}},\
			\bibinfo {pages} {050401} (\bibinfo {year} {2016})}\BibitemShut {NoStop}%
		\bibitem [{\citenamefont {Hlou{\v{s}}ek}\ \emph {et~al.}(2017)\citenamefont
			{Hlou{\v{s}}ek}, \citenamefont {Je{\v{z}}ek},\ and\ \citenamefont
			{Filip}}]{Filip}%
		\BibitemOpen
		\bibfield  {author} {\bibinfo {author} {\bibfnamefont {J.}~\bibnamefont
				{Hlou{\v{s}}ek}}, \bibinfo {author} {\bibfnamefont {M.}~\bibnamefont
				{Je{\v{z}}ek}},\ and\ \bibinfo {author} {\bibfnamefont {R.}~\bibnamefont
				{Filip}},\ }\bibfield  {title} {\bibinfo {title} {Work and information from
				thermal states after subtraction of energy quanta},\ }\href
		{https://doi.org/10.1038/s41598-017-13502-0} {\bibfield  {journal} {\bibinfo
				{journal} {Scientific Reports}\ }\textbf {\bibinfo {volume} {7}},\ \bibinfo
			{pages} {13046} (\bibinfo {year} {2017})}\BibitemShut {NoStop}%
		\bibitem [{\citenamefont {Shu}\ \emph {et~al.}(2017)\citenamefont {Shu},
			\citenamefont {Dai},\ and\ \citenamefont {Scarani}}]{Scarani}%
		\BibitemOpen
		\bibfield  {author} {\bibinfo {author} {\bibfnamefont {A.}~\bibnamefont
				{Shu}}, \bibinfo {author} {\bibfnamefont {J.}~\bibnamefont {Dai}},\ and\
			\bibinfo {author} {\bibfnamefont {V.}~\bibnamefont {Scarani}},\ }\bibfield
		{title} {\bibinfo {title} {Power of an optical maxwell's demon in the
				presence of photon-number correlations},\ }\href
		{https://doi.org/10.1103/PhysRevA.95.022123} {\bibfield  {journal} {\bibinfo
				{journal} {Phys. Rev. A}\ }\textbf {\bibinfo {volume} {95}},\ \bibinfo
			{pages} {022123} (\bibinfo {year} {2017})}\BibitemShut {NoStop}%
		\bibitem [{\citenamefont {Zanin}\ \emph {et~al.}(2022)\citenamefont {Zanin},
			\citenamefont {Antesberger}, \citenamefont {Jacquet}, \citenamefont
			{Ribeiro}, \citenamefont {Rozema},\ and\ \citenamefont {Walther}}]{Zanin}%
		\BibitemOpen
		\bibfield  {author} {\bibinfo {author} {\bibfnamefont {G.~L.}\ \bibnamefont
				{Zanin}}, \bibinfo {author} {\bibfnamefont {M.}~\bibnamefont {Antesberger}},
			\bibinfo {author} {\bibfnamefont {M.~J.}\ \bibnamefont {Jacquet}}, \bibinfo
			{author} {\bibfnamefont {P.~H.~S.}\ \bibnamefont {Ribeiro}}, \bibinfo
			{author} {\bibfnamefont {L.~A.}\ \bibnamefont {Rozema}},\ and\ \bibinfo
			{author} {\bibfnamefont {P.}~\bibnamefont {Walther}},\ }\bibfield  {title}
		{\bibinfo {title} {Enhanced {P}hotonic {M}axwell's {D}emon with {C}orrelated
				{B}aths},\ }\href {https://doi.org/10.22331/q-2022-09-20-810} {\bibfield
			{journal} {\bibinfo  {journal} {{Quantum}}\ }\textbf {\bibinfo {volume}
				{6}},\ \bibinfo {pages} {810} (\bibinfo {year} {2022})}\BibitemShut {NoStop}%
		\bibitem [{\citenamefont {Zhang}(2021)}]{Zhang}%
		\BibitemOpen
		\bibfield  {author} {\bibinfo {author} {\bibfnamefont {S.}~\bibnamefont
				{Zhang}},\ }\bibfield  {title} {\bibinfo {title} {Optical maxwell's demon
				phase space theory and its efficiency improvement to 90 
				squeezing},\ }in\ \href {https://doi.org/10.1109/WCSP52459.2021.9613393}
		{\emph {\bibinfo {booktitle} {2021 13th International Conference on Wireless
					Communications and Signal Processing (WCSP)}}}\ (\bibinfo {year} {2021})\
		pp.\ \bibinfo {pages} {1--5}\BibitemShut {NoStop}%
		\bibitem [{\citenamefont {Tatsi}\ \emph {et~al.}(2022)\citenamefont {Tatsi},
			\citenamefont {Canning}, \citenamefont {Zanforlin}, \citenamefont
			{Mazzarella}, \citenamefont {Jeffers},\ and\ \citenamefont
			{Buller}}]{Buller}%
		\BibitemOpen
		\bibfield  {author} {\bibinfo {author} {\bibfnamefont {G.}~\bibnamefont
				{Tatsi}}, \bibinfo {author} {\bibfnamefont {D.~W.}\ \bibnamefont {Canning}},
			\bibinfo {author} {\bibfnamefont {U.}~\bibnamefont {Zanforlin}}, \bibinfo
			{author} {\bibfnamefont {L.}~\bibnamefont {Mazzarella}}, \bibinfo {author}
			{\bibfnamefont {J.}~\bibnamefont {Jeffers}},\ and\ \bibinfo {author}
			{\bibfnamefont {G.~S.}\ \bibnamefont {Buller}},\ }\bibfield  {title}
		{\bibinfo {title} {Manipulating thermal light via displaced-photon
				subtraction},\ }\href {https://doi.org/10.1103/PhysRevA.105.053701}
		{\bibfield  {journal} {\bibinfo  {journal} {Phys. Rev. A}\ }\textbf {\bibinfo
				{volume} {105}},\ \bibinfo {pages} {053701} (\bibinfo {year}
			{2022})}\BibitemShut {NoStop}%
		\bibitem [{\citenamefont {Olivares}(2012)}]{olivares-2012}%
		\BibitemOpen
		\bibfield  {author} {\bibinfo {author} {\bibfnamefont {S.}~\bibnamefont
				{Olivares}},\ }\bibfield  {title} {\bibinfo {title} {Quantum optics in the
				phase space},\ }\href {https://doi.org/10.1140/epjst/e2012-01532-4}
		{\bibfield  {journal} {\bibinfo  {journal} {The European Physical Journal
					Special Topics}\ }\textbf {\bibinfo {volume} {203}},\ \bibinfo {pages} {3}
			(\bibinfo {year} {2012})}\BibitemShut {NoStop}%
		\bibitem [{\citenamefont {Barnett}\ and\ \citenamefont
			{Radmore}(2002)}]{Barnett}%
		\BibitemOpen
		\bibfield  {author} {\bibinfo {author} {\bibfnamefont {S.~M.}\ \bibnamefont
				{Barnett}}\ and\ \bibinfo {author} {\bibfnamefont {P.~M.}\ \bibnamefont
				{Radmore}},\ }\href
		{https://doi.org/10.1093/acprof:oso/9780198563617.001.0001} {\emph {\bibinfo
				{title} {{Methods in Theoretical Quantum Optics}}}}\ (\bibinfo  {publisher}
		{Oxford University Press, Oxford},\ \bibinfo {year} {2002})\BibitemShut
		{NoStop}%
		\bibitem [{\citenamefont {Lvovsky}\ and\ \citenamefont
			{Mlynek}(2002)}]{catalysisprl2002}%
		\BibitemOpen
		\bibfield  {author} {\bibinfo {author} {\bibfnamefont {A.~I.}\ \bibnamefont
				{Lvovsky}}\ and\ \bibinfo {author} {\bibfnamefont {J.}~\bibnamefont
				{Mlynek}},\ }\bibfield  {title} {\bibinfo {title} {Quantum-optical catalysis:
				Generating nonclassical states of light by means of linear optics},\ }\href
		{https://doi.org/10.1103/PhysRevLett.88.250401} {\bibfield  {journal}
			{\bibinfo  {journal} {Phys. Rev. Lett.}\ }\textbf {\bibinfo {volume} {88}},\
			\bibinfo {pages} {250401} (\bibinfo {year} {2002})}\BibitemShut {NoStop}%
		\bibitem [{\citenamefont {Ulanov}\ \emph
			{et~al.}(2015{\natexlab{a}})\citenamefont {Ulanov}, \citenamefont {Fedorov},
			\citenamefont {Pushkina}, \citenamefont {Kurochkin}, \citenamefont {Ralph},\
			and\ \citenamefont {Lvovsky}}]{catalysisnature2015}%
		\BibitemOpen
		\bibfield  {author} {\bibinfo {author} {\bibfnamefont {A.~E.}\ \bibnamefont
				{Ulanov}}, \bibinfo {author} {\bibfnamefont {I.~A.}\ \bibnamefont {Fedorov}},
			\bibinfo {author} {\bibfnamefont {A.~A.}\ \bibnamefont {Pushkina}}, \bibinfo
			{author} {\bibfnamefont {Y.~V.}\ \bibnamefont {Kurochkin}}, \bibinfo {author}
			{\bibfnamefont {T.~C.}\ \bibnamefont {Ralph}},\ and\ \bibinfo {author}
			{\bibfnamefont {A.~I.}\ \bibnamefont {Lvovsky}},\ }\bibfield  {title}
		{\bibinfo {title} {Undoing the effect of loss on quantum entanglement},\
		}\href {https://doi.org/10.1038/nphoton.2015.195} {\bibfield  {journal}
			{\bibinfo  {journal} {Nature Photonics}\ }\textbf {\bibinfo {volume} {9}},\
			\bibinfo {pages} {764} (\bibinfo {year} {2015}{\natexlab{a}})}\BibitemShut
		{NoStop}%
		\bibitem [{\citenamefont {Nunn}\ \emph {et~al.}(2022)\citenamefont {Nunn},
			\citenamefont {Franson},\ and\ \citenamefont {Pittman}}]{sps22}%
		\BibitemOpen
		\bibfield  {author} {\bibinfo {author} {\bibfnamefont {C.~M.}\ \bibnamefont
				{Nunn}}, \bibinfo {author} {\bibfnamefont {J.~D.}\ \bibnamefont {Franson}},\
			and\ \bibinfo {author} {\bibfnamefont {T.~B.}\ \bibnamefont {Pittman}},\
		}\bibfield  {title} {\bibinfo {title} {Modifying quantum optical states by
				zero-photon subtraction},\ }\href
		{https://doi.org/10.1103/PhysRevA.105.033702} {\bibfield  {journal} {\bibinfo
				{journal} {Phys. Rev. A}\ }\textbf {\bibinfo {volume} {105}},\ \bibinfo
			{pages} {033702} (\bibinfo {year} {2022})}\BibitemShut {NoStop}%
		\bibitem [{\citenamefont {Nunn}\ \emph {et~al.}(2023)\citenamefont {Nunn},
			\citenamefont {Shringarpure},\ and\ \citenamefont {Pittman}}]{sps23}%
		\BibitemOpen
		\bibfield  {author} {\bibinfo {author} {\bibfnamefont {C.~M.}\ \bibnamefont
				{Nunn}}, \bibinfo {author} {\bibfnamefont {S.~U.}\ \bibnamefont
				{Shringarpure}},\ and\ \bibinfo {author} {\bibfnamefont {T.~B.}\ \bibnamefont
				{Pittman}},\ }\bibfield  {title} {\bibinfo {title} {Transforming photon
				statistics through zero-photon subtraction},\ }\href
		{https://doi.org/10.1103/PhysRevA.107.043711} {\bibfield  {journal} {\bibinfo
				{journal} {Phys. Rev. A}\ }\textbf {\bibinfo {volume} {107}},\ \bibinfo
			{pages} {043711} (\bibinfo {year} {2023})}\BibitemShut {NoStop}%
		\bibitem [{\citenamefont {Zhong}\ \emph {et~al.}(2020)\citenamefont {Zhong},
			\citenamefont {Guo}, \citenamefont {Mao}, \citenamefont {Ye},\ and\
			\citenamefont {Huang}}]{ZPC}%
		\BibitemOpen
		\bibfield  {author} {\bibinfo {author} {\bibfnamefont {H.}~\bibnamefont
				{Zhong}}, \bibinfo {author} {\bibfnamefont {Y.}~\bibnamefont {Guo}}, \bibinfo
			{author} {\bibfnamefont {Y.}~\bibnamefont {Mao}}, \bibinfo {author}
			{\bibfnamefont {W.}~\bibnamefont {Ye}},\ and\ \bibinfo {author}
			{\bibfnamefont {D.}~\bibnamefont {Huang}},\ }\bibfield  {title} {\bibinfo
			{title} {Virtual zero-photon catalysis for improving continuous-variable
				quantum key distribution via gaussian post-selection},\ }\href
		{https://doi.org/10.1038/s41598-020-73379-4} {\bibfield  {journal} {\bibinfo
				{journal} {Scientific Reports}\ }\textbf {\bibinfo {volume} {10}},\ \bibinfo
			{pages} {17526} (\bibinfo {year} {2020})}\BibitemShut {NoStop}%
		\bibitem [{\citenamefont {Gagatsos}\ \emph {et~al.}(2014)\citenamefont
			{Gagatsos}, \citenamefont {Fiur\'a\ifmmode~\check{s}\else \v{s}\fi{}ek},
			\citenamefont {Zavatta}, \citenamefont {Bellini},\ and\ \citenamefont
			{Cerf}}]{Pra14}%
		\BibitemOpen
		\bibfield  {author} {\bibinfo {author} {\bibfnamefont {C.~N.}\ \bibnamefont
				{Gagatsos}}, \bibinfo {author} {\bibfnamefont {J.}~\bibnamefont
				{Fiur\'a\ifmmode~\check{s}\else \v{s}\fi{}ek}}, \bibinfo {author}
			{\bibfnamefont {A.}~\bibnamefont {Zavatta}}, \bibinfo {author} {\bibfnamefont
				{M.}~\bibnamefont {Bellini}},\ and\ \bibinfo {author} {\bibfnamefont {N.~J.}\
				\bibnamefont {Cerf}},\ }\bibfield  {title} {\bibinfo {title} {Heralded
				noiseless amplification and attenuation of non-gaussian states of light},\
		}\href {https://doi.org/10.1103/PhysRevA.89.062311} {\bibfield  {journal}
			{\bibinfo  {journal} {Phys. Rev. A}\ }\textbf {\bibinfo {volume} {89}},\
			\bibinfo {pages} {062311} (\bibinfo {year} {2014})}\BibitemShut {NoStop}%
		\bibitem [{\citenamefont {Thapliyal}\ \emph {et~al.}(2017)\citenamefont
			{Thapliyal}, \citenamefont {Samantray}, \citenamefont {Banerji},\ and\
			\citenamefont {Pathak}}]{THAPLIYAL20173178}%
		\BibitemOpen
		\bibfield  {author} {\bibinfo {author} {\bibfnamefont {K.}~\bibnamefont
				{Thapliyal}}, \bibinfo {author} {\bibfnamefont {N.~L.}\ \bibnamefont
				{Samantray}}, \bibinfo {author} {\bibfnamefont {J.}~\bibnamefont {Banerji}},\
			and\ \bibinfo {author} {\bibfnamefont {A.}~\bibnamefont {Pathak}},\
		}\bibfield  {title} {\bibinfo {title} {Comparison of lower- and higher-order
				nonclassicality in photon added and subtracted squeezed coherent states},\
		}\href {https://doi.org/https://doi.org/10.1016/j.physleta.2017.08.019}
		{\bibfield  {journal} {\bibinfo  {journal} {Physics Letters A}\ }\textbf
			{\bibinfo {volume} {381}},\ \bibinfo {pages} {3178} (\bibinfo {year}
			{2017})}\BibitemShut {NoStop}%
		\bibitem [{\citenamefont {Malpani}\ \emph {et~al.}(2019)\citenamefont
			{Malpani}, \citenamefont {Alam}, \citenamefont {Thapliyal}, \citenamefont
			{Pathak}, \citenamefont {Narayanan},\ and\ \citenamefont {Banerjee}}]{Priya}%
		\BibitemOpen
		\bibfield  {author} {\bibinfo {author} {\bibfnamefont {P.}~\bibnamefont
				{Malpani}}, \bibinfo {author} {\bibfnamefont {N.}~\bibnamefont {Alam}},
			\bibinfo {author} {\bibfnamefont {K.}~\bibnamefont {Thapliyal}}, \bibinfo
			{author} {\bibfnamefont {A.}~\bibnamefont {Pathak}}, \bibinfo {author}
			{\bibfnamefont {V.}~\bibnamefont {Narayanan}},\ and\ \bibinfo {author}
			{\bibfnamefont {S.}~\bibnamefont {Banerjee}},\ }\bibfield  {title} {\bibinfo
			{title} {Lower- and higher-order nonclassical properties of photon added and
				subtracted displaced fock states},\ }\href
		{https://doi.org/https://doi.org/10.1002/andp.201800318} {\bibfield
			{journal} {\bibinfo  {journal} {Annalen der Physik}\ }\textbf {\bibinfo
				{volume} {531}},\ \bibinfo {pages} {1800318} (\bibinfo {year}
			{2019})}\BibitemShut {NoStop}%
		\bibitem [{\citenamefont {Mandel}(1979)}]{Mandel:79}%
		\BibitemOpen
		\bibfield  {author} {\bibinfo {author} {\bibfnamefont {L.}~\bibnamefont
				{Mandel}},\ }\bibfield  {title} {\bibinfo {title} {Sub-poissonian photon
				statistics in resonance fluorescence},\ }\href
		{https://doi.org/10.1364/OL.4.000205} {\bibfield  {journal} {\bibinfo
				{journal} {Opt. Lett.}\ }\textbf {\bibinfo {volume} {4}},\ \bibinfo {pages}
			{205} (\bibinfo {year} {1979})}\BibitemShut {NoStop}%
		\bibitem [{\citenamefont {Biswas}\ and\ \citenamefont
			{Agarwal}(2007)}]{Biswas}%
		\BibitemOpen
		\bibfield  {author} {\bibinfo {author} {\bibfnamefont {A.}~\bibnamefont
				{Biswas}}\ and\ \bibinfo {author} {\bibfnamefont {G.~S.}\ \bibnamefont
				{Agarwal}},\ }\bibfield  {title} {\bibinfo {title} {Nonclassicality and
				decoherence of photon-subtracted squeezed states},\ }\href
		{https://doi.org/10.1103/PhysRevA.75.032104} {\bibfield  {journal} {\bibinfo
				{journal} {Phys. Rev. A}\ }\textbf {\bibinfo {volume} {75}},\ \bibinfo
			{pages} {032104} (\bibinfo {year} {2007})}\BibitemShut {NoStop}%
		\bibitem [{\citenamefont {Lee}(1990)}]{antibunching}%
		\BibitemOpen
		\bibfield  {author} {\bibinfo {author} {\bibfnamefont {C.~T.}\ \bibnamefont
				{Lee}},\ }\bibfield  {title} {\bibinfo {title} {Many-photon antibunching in
				generalized pair coherent states},\ }\href
		{https://doi.org/10.1103/PhysRevA.41.1569} {\bibfield  {journal} {\bibinfo
				{journal} {Phys. Rev. A}\ }\textbf {\bibinfo {volume} {41}},\ \bibinfo
			{pages} {1569} (\bibinfo {year} {1990})}\BibitemShut {NoStop}%
		\bibitem [{\citenamefont {Genoni}\ \emph {et~al.}(2008)\citenamefont {Genoni},
			\citenamefont {Paris},\ and\ \citenamefont {Banaszek}}]{non-G}%
		\BibitemOpen
		\bibfield  {author} {\bibinfo {author} {\bibfnamefont {M.~G.}\ \bibnamefont
				{Genoni}}, \bibinfo {author} {\bibfnamefont {M.~G.~A.}\ \bibnamefont
				{Paris}},\ and\ \bibinfo {author} {\bibfnamefont {K.}~\bibnamefont
				{Banaszek}},\ }\bibfield  {title} {\bibinfo {title} {Quantifying the
				non-gaussian character of a quantum state by quantum relative entropy},\
		}\href {https://doi.org/10.1103/PhysRevA.78.060303} {\bibfield  {journal}
			{\bibinfo  {journal} {Phys. Rev. A}\ }\textbf {\bibinfo {volume} {78}},\
			\bibinfo {pages} {060303} (\bibinfo {year} {2008})}\BibitemShut {NoStop}%
		\bibitem [{\citenamefont {Hu}\ \emph {et~al.}(2010)\citenamefont {Hu},
			\citenamefont {Xu}, \citenamefont {Wang},\ and\ \citenamefont
			{Xu}}]{pra2010}%
		\BibitemOpen
		\bibfield  {author} {\bibinfo {author} {\bibfnamefont {L.-y.}\ \bibnamefont
				{Hu}}, \bibinfo {author} {\bibfnamefont {X.-x.}\ \bibnamefont {Xu}}, \bibinfo
			{author} {\bibfnamefont {Z.-s.}\ \bibnamefont {Wang}},\ and\ \bibinfo
			{author} {\bibfnamefont {X.-f.}\ \bibnamefont {Xu}},\ }\bibfield  {title}
		{\bibinfo {title} {Photon-subtracted squeezed thermal state: Nonclassicality
				and decoherence},\ }\href {https://doi.org/10.1103/PhysRevA.82.043842}
		{\bibfield  {journal} {\bibinfo  {journal} {Phys. Rev. A}\ }\textbf {\bibinfo
				{volume} {82}},\ \bibinfo {pages} {043842} (\bibinfo {year}
			{2010})}\BibitemShut {NoStop}%
		\bibitem [{\citenamefont {Ulanov}\ \emph
			{et~al.}(2015{\natexlab{b}})\citenamefont {Ulanov}, \citenamefont {Fedorov},
			\citenamefont {Pushkina}, \citenamefont {Kurochkin}, \citenamefont {Ralph},\
			and\ \citenamefont {Lvovsky}}]{Ulanov2015}%
		\BibitemOpen
		\bibfield  {author} {\bibinfo {author} {\bibfnamefont {A.~E.}\ \bibnamefont
				{Ulanov}}, \bibinfo {author} {\bibfnamefont {I.~A.}\ \bibnamefont {Fedorov}},
			\bibinfo {author} {\bibfnamefont {A.~A.}\ \bibnamefont {Pushkina}}, \bibinfo
			{author} {\bibfnamefont {Y.~V.}\ \bibnamefont {Kurochkin}}, \bibinfo {author}
			{\bibfnamefont {T.~C.}\ \bibnamefont {Ralph}},\ and\ \bibinfo {author}
			{\bibfnamefont {A.~I.}\ \bibnamefont {Lvovsky}},\ }\bibfield  {title}
		{\bibinfo {title} {Undoing the effect of loss on quantum entanglement},\
		}\href {https://doi.org/10.1038/nphoton.2015.195} {\bibfield  {journal}
			{\bibinfo  {journal} {Nature Photonics}\ }\textbf {\bibinfo {volume} {9}},\
			\bibinfo {pages} {764} (\bibinfo {year} {2015}{\natexlab{b}})}\BibitemShut
		{NoStop}%
		\bibitem [{\citenamefont {Meena}\ and\ \citenamefont
			{Banerjee}(2023)}]{Meena2023}%
		\BibitemOpen
		\bibfield  {author} {\bibinfo {author} {\bibfnamefont {R.}~\bibnamefont
				{Meena}}\ and\ \bibinfo {author} {\bibfnamefont {S.}~\bibnamefont
				{Banerjee}},\ }\bibfield  {title} {\bibinfo {title} {Characterization of
				quantumness of non-gaussian states under the influence of gaussian channel},\
		}\href {https://doi.org/10.1007/s11128-023-04037-7} {\bibfield  {journal}
			{\bibinfo  {journal} {Quantum Information Processing}\ }\textbf {\bibinfo
				{volume} {22}},\ \bibinfo {pages} {298} (\bibinfo {year} {2023})}\BibitemShut
		{NoStop}%
	\end{thebibliography}
	%
	
\end{document}